\documentclass[11pt]{iopart}

\expandafter\let\csname equation*\endcsname\relax
\expandafter\let\csname endequation*\endcsname\relax
\usepackage{graphicx}
\usepackage{amssymb}
\usepackage{amsmath}
\usepackage{amsthm}
\usepackage{cite}
\usepackage{hyperref}
\usepackage{epstopdf}
\usepackage{tikz}
\usepackage{caption}
\usepackage{subcaption}
\usepackage{hyperref}
\usepackage{float}
\usepackage{enumitem}
\usepackage{calc}

\theoremstyle{remark}

\usepackage[font=small, labelfont=bf, width=0.8\textwidth]{caption}

\def\be{\begin{equation}}
\def\ee{\end{equation}}

\def\bfz{{\bf z}}

\def\bfJ{{\bf J}}
\def\bfrho{\boldsymbol{\rho}}

\makeatletter
\newcommand*{\textoverline}[1]{$\overline{\hbox{#1}}\m@th$}
\makeatother

\begin{document}


\title[Hydrodynamic behavior of the  2--TASEP]  
{Hydrodynamic behavior of the  2--TASEP}

\author{Luigi~Cantini}
\address{Laboratoire de Physique Th\'eorique et Mod\'elisation
  (CNRS UMR 8089), 
CY Cergy Paris Universit\'e, F-95302 Cergy-Pontoise, France}
\eads{luigi.cantini@cyu.fr}

\author{Ali~Zahra}
\address{Laboratoire de Physique Th\'eorique et Mod\'elisation
  (CNRS UMR 8089), 
CY Cergy Paris Universit\'e, F-95302 Cergy-Pontoise, France}
\eads{ali.zahra@cyu.fr}

\begin{abstract}
We address the question of the large scale or hydrodynamic 
behavior of a 2-species generalization of TASEP (2--TASEP), consisting of two kinds of particles, moving in opposite directions and swapping their positions. 
We compute the rarefaction and shock solutions of the hydrodynamic equations of the model, showing that these equations form a Temple class system.
We solve completely the Riemann problem and compare the theoretical prediction to Monte Carlo simulations.
\end{abstract}

\maketitle 



\section{Introduction}

The asymmetric simple exclusion process (ASEP) is a minimal model of transport in (quasi) one--dimensional systems. It consists of particles which occupy the sites of a one dimensional lattice with only one particle allowed on each lattice site.
These particles hop under the effect of an external driving force which breaks detailed balance and creates a stationary current. 
This model was introduced independently in the late 60s in biology (to model translation in
protein synthesis) \cite{macdonald1968kinetics} and in probability \cite{spitzer1970interaction} and afterwards it has found a wide spectrum of applications, ranging from theoretical and experimental studies of biophysical transport \cite{chou2011non} to 
%
%
%
%
modeling  traffic  flow  \cite{chowdhury2000statistical,evans1996bose}. 
As soon one considers models which are more suited for physical/biological systems, one will encounter variants of ASEPs containing localized or mobile defects and several species of particles, which have different behaviors.  As a result, typically these models are not exactly solvable and even 
for some of the most basic questions, like the study of large scale behavior of the system (which in the case of ASEP is known to be described by the Burgers equation \cite{rost1981non,benassi1987hydrodynamical,
rezakhanlou1991hydrodynamic,kipnis1998scaling}), approximations schemes like mean--field are  necessary  for  further  progress. 

In this paper we address the question of the large scale or hydrodynamic 
behavior of an exactly solvable multispecies generalization of ASEP, consisting of two kinds of particles, $\bullet$--particles and $\circ$--particles, moving in opposite directions. One can think of them as opposite charged particles moving under the influence of an external electric field or as cars moving on two opposite lanes.
Each site of a one--dimensional lattice is either empty or occupied by one of the two kinds of particles and empty sites can
be  treated as a third species of particles, the $\ast$--particles.  
%
%
In continuous time, a $\bullet$--particle jumps forward on empty sites  with rate $\beta$, while a white $\circ$--particle jumps backward on empty sites with rate $\alpha$. On top of this, an adjacent pair $\bullet\circ$ swaps to $\circ\bullet$ with rate $1$. 
\be
\begin{split}
\bullet\,\ast& ~ \rightarrow ~ \ast\,\bullet\quad\text{rate}\quad \beta\\
\ast\,\circ& ~ \rightarrow ~ \circ\,\ast\quad\text{rate}\quad \alpha\\
\bullet\,\circ& ~ \rightarrow ~ \circ\,\bullet\quad\text{rate}\quad 1
\end{split}
\ee
  This model has appeared in the literature under different names. It has been first considerd in \cite{derrida1996statphys,mallick1996shocks}, where the stationary measure on a finite periodic lattice was written in a matrix product ansatz form \cite{derrida1993exact,blythe2007nonequilibrium}. It is also a particular case ($q=0$) of the so called AHR model 
\cite{arndt1998spontaneous,arndt1999spontaneous,rajewsky2000spatial},  in which the swap $\circ\bullet\rightarrow \bullet\circ$ is allowed with rate $q$.
Being this model a natural 2--species generalization of TASEP we shall call it $2$--TASEP.

It turns out that the 2--TASEP is Yang--Baxter integrable
\cite{cantini2008algebraic} for arbitrary values of the parameters $
\alpha$ and $\beta$. It belongs indeed to a larger family of integrable multispecies exclusion processes introduced in \cite{cantini2016inhomogenous}. Bethe ansatz techniques can be used to solve  exactly 
for the long time limit behaviour of the generating function of the  
currents \cite{derrida1999bethe,cantini2008algebraic}. 
More recently, 
in the case $\alpha+\beta=1$, the transition probabilities as well as the 
joint current distribution for some specific initial distribution of a 
finite number of $\bullet$--particles and vacancies have been obtained 
\cite{chen2018exact,chen2021limiting}, and an asymptotic analysis of 
these results has allowed to prove that the joint current distribution is 
given by a product of a Gaussian and a GUE Tracy-Widom distribution in the 
long time limit, as predicted by non--linear fluctuating hydrodynamics
 \cite{van2012exact,spohn2014nonlinear,ferrari2013coupled}.
When $\alpha+\beta=1$ the stationary measure factorizes and the currents 
have a simple expression as function of the densities.
In \cite{fritz2004derivation,toth2005perturbation} the 
hydrodynamic limit of the 2--TASEP for $\alpha=\beta=\frac{1}{2}$ has 
been 
studied and proven to converge to the classical Leroux system of 
conservation laws \cite{leroux1978analyse,serre1988existence}.
The Leroux system is a notable example of a \emph{Temple class system} 
i.e. e a 2--components conservation law whose shock and rarefaction 
curves coincide  \cite{temple1983systems}. 
For arbitrary $\alpha$ and $\beta$ only numerical results based on  mean field approximation are available 
\cite{arndt2002spontaneous}.
 In the present paper we study the exact hydrodynamic equations of the 2--TASEP for arbitrary $\alpha$ and $\beta$. We compute their rarefaction solutions and their shocks, showing that the hydrodynamic equations of 2--TASEP are a Temple class system.

%

\vspace{.3cm}

 The paper is organized as follows. In Section \ref{sect:curr} we review and expand on results about the 2--TASEP currents obtained in \cite{cantini2008algebraic}. The core of the paper is Section \ref{sect:conserv} where the conservation laws are studied. We derive the rarefaction waves as well as the shock solutions and in Section
 \ref{sect-riemann} we solve  the Riemann problem. In the final Section \ref{sec:montecarlo} we compare the prediction of the hydrodynamic equations with Monte Carlo simulations.

\section{Currents}\label{sect:curr}

In this section we reproduce and expand the results of the analysis in \cite{cantini2008algebraic} in a convenient way, which makes manifest the symmetries of the model. 
In order to derive the dependence of the particle currents as function of 
the local densities, we consider our model on a periodic ring with fixed 
number of particles of each species. Call $M_i$ the number of particles of 
$i$-th species, they are related to $N$, the length of the ring, by $N=M_\bullet+M_\circ+M_\ast$.
Let the system evolve starting at time $t=0$ from an arbitrary fixed configuration and call $n_{i,j}(t)$ the number of swaps of consecutive ordered pairs of particles of type $i,j$ up to time $t$. This number increases by $+1$ each time two consecutive ordered  particles of species $i,j$ exchange their position $i,j\rightarrow j,i$. 
The average rate of swaps  $i,j\rightarrow j,i$ in the steady state is just given by $\lim_{t\rightarrow+\infty}\frac{1}{t}\mathbb{E}\left[  n_{i,j}(t)\right]$, irrespectively of the initial state. The particle currents in the steady state are hence given by 
\begin{equation}
J_i= \lim_{t\rightarrow+\infty}\frac{1}{Nt}\mathbb{E}\left[ \sum_{j\ne i} n_{i,j}(t)-n_{j,i}(t)\right]
\end{equation}
In our case, it is convenient to introduce the following quantity
\be 
\Phi(\nu_{\bullet,\circ},\nu_{\bullet,\ast},\nu_{\ast,\circ})=\lim_{t\rightarrow+\infty}\frac{1}{t}\mathbb{E}\left[\nu_{\bullet,\circ}\,  n_{\bullet,\circ}(t)  + \nu_{\bullet,\ast}\, 
 n_{\bullet,\ast}(t) 
+\nu_{\ast,\circ}\, n_{\ast,\circ}(t) \right].
\ee
The currents are obtained as specialization of $\Phi(\nu_{\bullet,\circ},\nu_{\bullet,\ast},\nu_{\ast,\circ})$
\be 
J_\bullet=\frac{\Phi(1,1,0)}{N},\quad J_\circ=	\frac{ \Phi(-1,0,-1)}{N},\quad J_\ast=\frac{\Phi(0,-1,1)}{N}.
\ee
IN \cite{cantini2008algebraic} the function $\Phi(\nu_{\bullet,\circ},\nu_{\bullet,\ast},\nu_{\ast,\circ})$  was show to be given by the solution of the following equation
\be\label{det-phi} 
\det G(\Phi(\nu_{\bullet,\circ},\nu_{\bullet,\ast},\nu_{\ast,\circ}),\nu_{\bullet,\circ},\nu_{\bullet,\ast},\nu_{\ast,\circ})=0.
\ee
where the matrix $G(\Phi,\nu_{\bullet,\circ},\nu_{\bullet,\ast},\nu_{\ast,\circ})$ is given by
\be 
G(\Phi,\nu_{\bullet,\circ},\nu_{\bullet,\ast},\nu_{\ast,\circ})=
\scalebox{.75}{$\left(
\begin{array}{ccc}
\Phi& F_\alpha[M_\circ,M_\bullet,M_\ast] & F_\beta[M_\bullet,M_\circ,M_\ast]\\
\nu_{\bullet,\circ}M_\bullet+\nu_{\ast,\circ}M_\ast& F_\alpha[M_\circ+1,M_\bullet,M_\ast] & -F_\beta[M_\bullet,M_\circ+1,M_\ast]\\
\nu_{\bullet,\circ}M_\circ+\nu_{\bullet,\ast}M_\ast &
-F_\alpha[M_\circ,M_\bullet+1,M_\ast]& F_\beta[M_\bullet+1,M_\circ,M_\ast]
\end{array}
\right)$}
\ee
with
\be
F_\gamma[a,b,c]:= \oint_0 \frac{dz}{2\pi i}
\frac{1}{z^{a}(z-1)^{b}(z-\gamma)^{c}}.
\ee
When one of the particle species
is \emph{strictly} absent (i.e. when one among $M_\bullet,M_\circ,M_\ast$ vanishes) the model reduces to a single species TASEP and it is not difficult to see that one of the currents vanishes, while the others boil down to the usual TASEP  current. On the other hand in the following we shall assume that at least one particle per species is present ($M_i\neq 0$) and we shall be mainly interested in the thermodynamic limit of these quantities as $N\rightarrow \infty$, with $\lim_{N\rightarrow\infty}\frac{M_i}{N}=\rho_i$. 
We shall see, as already found in \cite{mallick1996shocks,derrida1996statphys}, that the presence of even a single particle of a given species (i.e. an infinitesimally vanishing but not strictly zero density) can affect the macroscopic behavior of the system. 
With this in mind, we consider the limit 
$a\longrightarrow \infty$, with $b/a$  and $c/a$ fixed, of the function
$F_\gamma[a,b,c]$,
that behaves like\footnote{Here we are supposing $a,b,c,\gamma>0$. } 
$$
F_\gamma[a,b,c]\sim \frac{1}{z_\gamma^a(z_\gamma-1)^{b}(z_\gamma-\gamma)^{c}}
$$
where $z_\gamma$ is the zero  of the saddle point equation
$
\frac{a}{z}+\frac{b}{z-1}+\frac{c}{z-\gamma}=0
$,  belonging to the interval $[0,\min[1,\gamma]]$.
Applying this expression in eq.(\ref{det-phi})
 we get in the thermodynamic limit  
\be 
\lim_{N\rightarrow \infty}\frac{\Phi(\nu_{\bullet,\circ},\nu_{\bullet,\ast},\nu_{\ast,\circ})}{N}=(\nu_{\bullet,\circ}\rho_\bullet+\nu_{\ast,\circ}\rho_\ast)z_\alpha(1-z_\beta)+(\nu_{\bullet,\circ}\rho_\circ+\nu_{\bullet,\ast}\rho_\ast)z_\beta(1-z_\alpha)
\ee
where with $z_\alpha \in [0,\min(1,\alpha)]$ and $z_\beta \in [0,\min(1,\beta)]$ are solution of the saddle point equations
\begin{gather}\label{chang-var1}
\frac{\rho_\circ}{z_\alpha}+\frac{\rho_\bullet}{z_\alpha-1}+\frac{1-\rho_\circ-\rho_\bullet}{z_\alpha-\alpha}=0\\ \label{chang-var2}
\frac{\rho_\bullet}{z_\beta}+\frac{\rho_\circ}{z_\beta-1}+\frac{1-\rho_\circ-\rho_\bullet}{z_\beta-\beta}=0.
\end{gather}
The result for the currents then reads
\begin{gather}\label{0Jrz} 
J_\circ= z_\alpha(z_\beta-1)+\rho_\circ(z_\alpha-z_\beta)\\
\label{1Jrz}
J_\bullet= z_\beta(1-z_\alpha)+\rho_\bullet(z_\alpha-z_\beta)\\
\label{2Jrz}
J_\ast= \rho_\ast(z_\alpha-z_\beta)
\end{gather}
Notice that eqs.(\ref{chang-var1},\ref{chang-var2}) are invariant 
under exchange $\rho_\circ\leftrightarrow \rho_\bullet$, $\alpha
\leftrightarrow \beta$ and $z_\alpha\leftrightarrow z_\beta$. This 
implies as expected, that under 
exchange $\rho_\circ\leftrightarrow \rho_\bullet$ and $\alpha\leftrightarrow
\beta$   we have  
$J_\circ\leftrightarrow-J_\bullet$.
Let us finish this section by showing how some known results fit in the analysis here above.   
\begin{itemize}[leftmargin=0.5cm]
\item [$\Diamond$] $\beta=1$. In this case $\bullet$--particles dont distinguish $\circ$--particles from $\ast$--particles and so they behave just as particles in a single species TASEP. 
This is reflected in eq.(\ref{chang-var2}), where $\rho_\circ$ disappears 
and one finds $z_\beta=\rho_\bullet$, which replaced in eq.(\ref{1Jrz}) 
gives $J_\bullet=\rho_\bullet(1-\rho_\bullet)$. The case $\alpha=1$ is 
completely analogous: $z_\alpha=\rho_\circ$ and $J_\circ=\rho_\circ(\rho_\circ-1)$.  
\item [$\Diamond$] $\alpha+\beta=1$. In this case it is known  that the stationary measure takes a factorized form  \cite{rajewsky2000spatial}. At the level of the currents, we have indeed $J_\circ=-\rho_\circ(\rho_\bullet+\alpha \rho_\ast)$ and $J_\circ=\rho_\bullet(\rho_\circ+\beta \rho_\ast)$.
\end{itemize}

\subsection{The $z$ variables}

In our analysis the variables $\bfz=(z_\alpha, z_\beta)$ will play a prominent role, it is therefore important to work out their domain of definition $\mathcal{D}_z(\alpha,\beta)$ corresponding to the physical domain 
$\mathcal{D}$ in the variables $\bfrho=(\rho_\circ,\rho_\bullet)$, $\rho_\circ,\rho_\bullet\geq 0, \rho_\circ+\rho_\bullet\leq 1$.
%
%
%
First of all we have already seen above that $\bfz$ has to satisfy  $z_\alpha \in [0,\min(1,\alpha)]$ and $z_\beta \in [0,\min(1,\beta)]$. 
At fixed $\bfz$, the system of equations (\ref{chang-var1},\ref{chang-var2}) is just the crossing of two lines in the $\bfrho$ plane: $\ell_\alpha$ coming from eq.(\ref{chang-var1}) and $\ell_\beta$ coming from eq.(\ref{chang-var2}). So we have to determine under which conditions these lines cross inside $\mathcal{D}$.

\begin{center}
\begin{tikzpicture}[scale=1.4]
\draw [fill=white!95!blue] (0,0)--(2,0) node [below] {$1$}--(0,2)node [left] {$1$}--cycle;
\draw [thick,->] (-.5,0)--(3,0) node[below] {$\rho_\circ$};
\draw [thick,->] (0,-.5)--(0,3) node[left] {$\rho_\bullet$};
\draw[blue,domain=-.5:2.5,smooth,variable=\t]plot (\t,0.2*\t+1);
\draw (2.5,0.2*2.5+1) node[above] {$\ell_\beta$};
\draw[fill=black] (0,1) circle  (.03) node[above left] {$\frac{z_\beta}{\beta}$};
\draw[fill=black] (0.83,1.17) circle  (.03) node[below right] {\quad \scriptsize $(1-z_\beta,z_\beta)$};
\draw[red,domain=-.5:2.5,smooth,variable=\t]plot (-0.1*\t+0.5,\t);
\draw (-0.1*2.5+0.5,2.5) node[right] {$\ell_\alpha$};
\draw[fill=black] (0.5,0) circle  (.03) node[below right] {$\frac{z_\alpha}{\alpha}$};
\draw[fill=black] (0.33,1.67) circle  (.03) node[right] {\scriptsize $(z_\alpha,1-z_\alpha)$};
\end{tikzpicture}
\end{center}
The line $\ell_\alpha$ crosses the $\rho_\bullet=0$ axis at $\rho_\circ=z_\alpha/\alpha$ and the $\rho_\circ+\rho_\bullet=1$ line at $\rho_\circ=z_\alpha,\rho_\bullet=1-z_\alpha$. The straight line $\ell_\beta$ crosses the $\rho_\circ=0$ axis at $\rho_\bullet=z_\beta/\beta$ and the $\rho_\circ+\rho_\bullet=1$ line at $\rho_\circ=1-z_\beta,\rho_\bullet=z_\beta$. So given $z_\alpha \in [0,\min(1,\alpha)]$ and $z_\beta \in [0,\min(1,\beta)]$, the necessary and sufficient condition for the lines $\ell_\alpha$ and $\ell_\beta$ to cross inside $\mathcal{D}$ is that $z_\alpha+z_\beta\leq 1$.
This means the physical domain $\mathcal{D}_z(\alpha,\beta)$ in the plane $z_\alpha, z_\beta$  is defined by $0\leq z_\alpha \leq \min(1,\alpha),0\leq z_\beta \leq \min(1,\beta) $ and $z_\alpha+z_\beta\leq 1$. 
From this simple geometrical argument we also conclude that (at fixed $z_
\alpha$) $\rho_\bullet$ is an increasing function of $z_\beta$, while (at fixed 
$z_\beta$) $\rho_\circ$ is an increasing function of $z_\alpha$.

\vspace{.2cm}

\begin{figure}[h!]
\centering
	\begin{subfigure}{.6\linewidth}
	\begin{tikzpicture}
	\begin{scope}[scale = .4]
\fill [fill=white!95!blue] (0,0)--(0,6)--(6,0)--cycle;
	\draw[thick,->] (0,0) -- (7,0) node[below] {\scriptsize $\rho_\circ$};
	\draw[thick,->] (0,0) -- (0,7) node[left] {\scriptsize $\rho_\bullet$};
	\foreach \x in {1,2,...,8}
		 \draw [red] (\x/3,0) -- (\x*2/3, 6- \x*2/3);
	
	\foreach \y in {1,2,...,8}
		\draw [blue] (0,4*\y/9) -- (6-\y*2/3, \y*2/3);
	
	\draw [line width=0.7 mm, black!40!green] (0,6) -- (6,0);
	\draw [line width=0.7 mm, red ]  (3,0) -- (6,0);
	\draw [line width=0.7 mm, blue ] (0,4) -- (0,6);

	\draw [ultra thick](-3pt,4 cm) -- (3pt,4 cm) node[left] {\scriptsize $\frac{1}{\beta}$};
	\draw [ultra thick] (3 cm,3pt) -- (3 cm,-3pt) node[below] {\scriptsize $\frac{1}{\alpha}$};
	\draw [thick](2pt,6 cm) node[left] {\scriptsize $1$};
	\draw [thick](6 cm,-2pt) node[below] {\scriptsize $1$};
	\end{scope}
\begin{scope}[scale = .4, xshift=14cm]
	\fill [fill=white!95!blue] (0,0)--(0,6)--(6,0)--cycle;
	\draw[thick,->] (0,0) -- (7,0) node[below] {\scriptsize $z_{\alpha}$};
	\draw[thick,->] (0,0) -- (0,7) node[left] {\scriptsize $z_{\beta}$};
	
	\foreach \x in {1,2,...,8}
	\draw [red] (2*\x/3,0) -- (2*\x/3, 6- 2*\x/3);
	
	\foreach \y in {1,2,...,8}
	\draw [blue] (0,2*\y/3) -- (6-2*\y/3, 2*\y/3);

	\draw [line width=0.7 mm, black!40!green] (0,6) -- (6,0);

	\fill [blue]  (0,6) circle (1mm) ;
	\fill [red]  (6,0) circle (1mm) ;
	
	\draw [thick](2pt,6 cm) node[anchor=east] {\scriptsize $1$};
	\draw [thick](6 cm,-2pt) node[anchor=north] {\scriptsize $1$};
	
	\end{scope}
	\end{tikzpicture}
	\caption{$\alpha,\beta>1$}\label{domains-maps-a}
	\end{subfigure}
	\begin{subfigure}[b]{.6\linewidth}
	\begin{tikzpicture}
	\begin{scope}[scale = .4]
	\fill [fill=white!95!blue] (0,0)--(0,6)--(6,0)--cycle;

\draw[thick,->] (0,0) -- (7,0) node[below] {\scriptsize $\rho_\circ$};
\draw[thick,->] (0,0) -- (0,7) node[left] {\scriptsize $\rho_\bullet$};

\draw [ultra thick] (6*0.2,6-6*0.2)++(45:-3pt)--++(45:6pt) node [above right] {\scriptsize $(1-\beta,\beta)$};

\draw [ultra thick] (6-0.4*6,0.4*6)++(45:-3pt)--++(45:6pt) node [above right] {\scriptsize $(\alpha,1-\alpha)$};

\foreach \x in {1,2,...,8}
\draw [red] (\x*2/3,0) -- (0.6*\x*2/3, 6- 0.6*\x*2/3);

\foreach \y in {1,2,...,8}
\draw [blue] (0,\y*2/3) -- (6-0.8*\y*2/3, 0.8*\y*2/3);

\draw [line width=0.7 mm, black!40!green] (0,6) -- (6,0);

\draw[line width=0.7 mm, blue] (6*0.2,6-6*0.2) -- (0,6);
\draw [line width=0.7 mm, red] (6,0) -- (6-0.4*6,0.4*6);

	\draw [thick] (2pt,6 cm) node[left] {\scriptsize $1$};
\draw [thick] (6 cm,-2pt) node[below] {\scriptsize $1$};
\end{scope}

\begin{scope}[scale = .4, xshift=14cm]
	\fill [fill=white!95!blue] (0,0)--(0,0.8*6) -- (6-0.8*6,0.8*6) --(0.6*6,6-0.6*6)--(0.6*6,0)--cycle;

\draw[thick,->] (0,0) -- (7,0) node[below] {\scriptsize $z_{\alpha}$};
\draw[thick,->] (0,0) -- (0,7) node[left] {\scriptsize $z_{\beta}$};

\foreach \x in {3,4,...,8}
\draw [red] (0.6*\x*2/3,0) -- (0.6*\x*2/3, 6- 0.6*\x*2/3);

\foreach \x in {1,2}
\draw [red] (0.6*\x*2/3,0) -- (0.6*\x*2/3, 6*0.8);

\foreach \y in {1,2,3,4}
	\draw [blue] (0,0.8*\y*2/3) -- (6*0.6, 0.8*\y*2/3);
\foreach \y in {5,6,7,8}
\draw [blue] (0,0.8*\y*2/3) -- (6-0.8*\y*2/3, 0.8*\y*2/3);

\draw [line width=0.7 mm, red] (0.6*6,0) -- (0.6*6,6-0.6*6);
\draw [line width=0.7 mm, blue] (0,0.8*6) -- (6-0.8*6,0.8*6);

\draw [line width=0.7 mm, black!40!green] (6-0.8*6,0.8*6) -- (0.6*6,6-0.6*6);

\draw (6*0.6 cm,-1pt) node[anchor=north] {\scriptsize $\alpha$};

\draw (-1pt,6*0.8 cm) node[anchor=east] {\scriptsize $\beta$};

	\draw [thick](-2pt,6 cm) -- (2pt,6 cm) node[anchor=east] {\scriptsize $1$};
\draw [thick] (6 cm,2pt) -- (6 cm,-2pt) node[anchor=north] {\scriptsize $1$};

\end{scope}

\end{tikzpicture}
		\caption{$\alpha,\beta <1$, $\alpha+\beta>1$}\label{domains-maps-b}

\end{subfigure}
	\begin{subfigure}[b]{.6\linewidth}
	\begin{tikzpicture}
	\begin{scope}[scale = .4]
	\fill [fill=white!95!blue] (0,0)--(0,6)--(6,0)--cycle;
	\draw[thick,->] (0,0) -- (7,0) node[below] {\scriptsize $\rho_\circ$};
\draw[thick,->] (0,0) -- (0,7) node[left] {\scriptsize $\rho_\bullet$};

\foreach \x in {1,2,3,4,5}
\draw [red] (\x,0) -- (0.3*\x, 6- 0.3*\x);

\foreach \y in {1,2,3,4,5}
\draw [blue] (0,\y) -- (6-0.4*\y, 0.4*\y);

\draw [line width=0.7 mm, black!40!purple] (6*0.2,6-6*0.2) -- (6-0.4*6,0.4*6);
\draw[line width=0.7 mm, blue] (6*0.3,6-6*0.3) -- (0,6);
\draw [line width=0.7 mm, red] (6,0) -- (6-0.4*6,0.4*6);

	\draw [thick](2pt,6 cm) node[left] {\scriptsize $1$};
	\draw [thick](6 cm,-2pt) node[below] {\scriptsize $1$};
\end{scope}

\begin{scope}[scale = .4, xshift=14cm]
\fill [fill=white!95!blue] (0,0)--(0,6*0.4) -- (6*0.3,6*0.4)--(6*0.3,0)--cycle;
\draw[thick,->] (0,0) -- (7,0) node[anchor=north west] {\scriptsize $z_{\alpha}$};
\draw[thick,->] (0,0) -- (0,7) node[anchor=south east] {\scriptsize $z_{\beta}$};

\foreach \x in {1,2,3,4,5}
\draw [red] (0.3*\x,0) -- (0.3*\x, 6*0.4);

\foreach \y in {1,2,3,4,5}
\draw [blue] (0,0.4*\y) -- (0.3*6, 0.4*\y);

\draw [line width=0.7 mm, blue]  (0,6*0.4) -- (6*0.3,6*0.4);
\draw [line width=0.7 mm, red]  (6*0.3,0) -- (6*0.3,6*0.4);

\fill [black!40!purple]  (6*0.3,6*0.4) circle (1mm) ;

\draw  (6*0.3 cm,-1pt) node[anchor=north] {\scriptsize $\alpha$};

\draw (-1pt,6*0.4 cm) node[anchor=east] {\scriptsize $\beta$};

	\draw [thick](-2pt,6 cm) -- (2pt,6 cm) node[anchor=east] {\scriptsize $1$};
\draw [thick] (6 cm,2pt) -- (6 cm,-2pt) node[anchor=north] {\scriptsize $1$};

\end{scope}
	\end{tikzpicture}
		\caption{$\alpha+\beta<1$}\label{domains-maps-c}
	\end{subfigure}
		\caption{On the left the physical domain $\mathcal{D}$ in the densities plane, on the right the corresponding domain $\mathcal{D}_z$ in the $\bfz$ variables plane.}\label{domains-maps}
	
\end{figure}
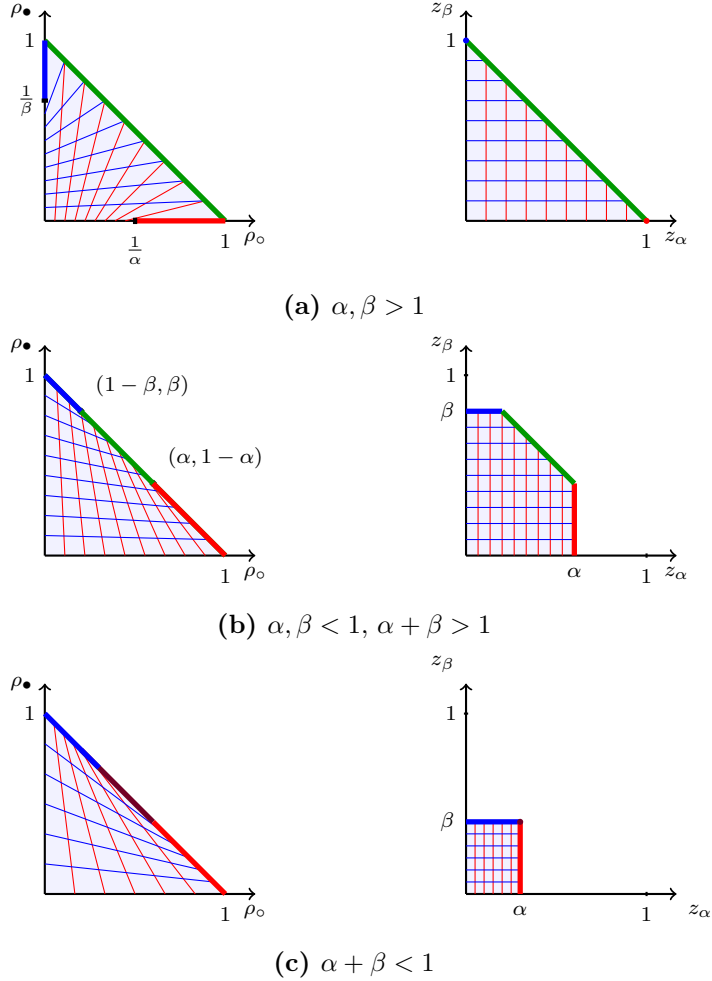

In Figures \ref{domains-maps-a}--\ref{domains-maps-c}  we have reported 
on the left the physical domain $\mathcal{D}$ in the densities plane and 
on the right the corresponding domain $\mathcal{D}_z$ in the $\bfz$ 
variables plane. 
Notice that in figure \ref{domains-maps-a} $(\alpha,\beta>1 )$ the thick 
red segment on the $\rho_\circ$ axis is mapped to the point $(z_
\alpha=1,z_\beta=0)$ and the thick blue segment on the $\rho_\bullet$ 
axis is mapped to the point  $(z_\alpha=0,z_\beta=1)$. 
In figure \ref{domains-maps-c}
 $(\alpha+\beta<1)$: the overlap of the red and blue segments on the boundary $\rho_\ast=0$ is mapped to the point  $(z_\alpha=\alpha,z_\beta=\beta)$.

This means that the mapping $\bfrho\rightarrow \bfz$ can be singular on the boundary of the physical domain, where at least one the densities vanishes.
Let's analyze the different possibilities and work out the portion of the boundary where the mapping is singular. 
\begin{itemize}[leftmargin=0cm] 
\item []\fbox{$\rho_\circ\rightarrow 0$} In this case from 
eq.(\ref{chang-var1}) we deduce that $z_\alpha= 0$ while the two 
solutions of eq.(\ref{chang-var2}) are $z_\beta=1,z_\beta=
\rho_\bullet\beta$ and we have to retain the smallest one. If $\beta\leq 
1$ then on the $\rho_\circ=0$ axis the map is $1$-to-$1$ and there are no 
singularities, on the other hand, if $\beta>1$ then all the points $\rho_
\bullet\geq \beta^{-1}$ are mapped to the same point $(z_\alpha=0,z_
\beta=1)$.
\item []\fbox{$\rho_\bullet\rightarrow 0$} This case is treated similarly to the previous one: we have $z_\beta= 0$ and the two solutions of eq.(\ref{chang-var1}) are $z_\alpha=1,z_\alpha=\rho_\circ\alpha$. If $\alpha \leq 1$ then on the $\rho_\bullet=0$ axis the map is $1$-to-$1$ and there are no singularities, on the other hand, if $\alpha>1$ then all the points $\rho_\circ\geq \alpha^{-1}$ are mapped to the same point $(z_\alpha=1,z_\beta=0)$.
\item []\fbox{$\rho_\ast\rightarrow 0$} Eqs.(\ref{chang-var1},
\ref{chang-var2}) have solutions, $z_\alpha=\alpha, z_\alpha=\rho_\circ$, 
$z_\beta=\beta, z_\beta=1-\rho_\circ$. Whenever $\alpha+\beta\leq 1$, all the points on the line 
$\rho_\circ+\rho_\bullet=1$ such that $\alpha\leq \rho_\circ\leq 1-\beta$ are 
mapped to the single point $z_\alpha=\alpha, z_\beta=\beta$. 
\end{itemize}

\subsection{Behaviour at the boundary of the physical domain}
The singularities of the mapping $\bfrho\rightarrow \bfz$ reflect some important features of the model. Let's consider  the currents
of non zero density particles at the boundary of $\mathcal{D}$

\noindent
\fbox{$\rho_\circ\rightarrow 0$} We have for the current 
\be\label{singul1}
J_\bullet(\rho_\bullet) =  \left\{
\begin{array}{ll}
\beta \rho_\bullet(1-\rho_\bullet) & 0\leq \rho_\bullet\leq \beta^{-1}\\
(1-\rho_\bullet) & \beta^{-1}\leq \rho_\bullet\leq 1.
\end{array}
\right.
\ee

\noindent
 \fbox{$\rho_\bullet\rightarrow 0$} We have for the current
\be\label{singul0}
J_\circ(\rho_\circ) =  \left\{
\begin{array}{ll}
-\alpha \rho_\circ(1-\rho_\circ) & 0\leq \rho_\circ\leq \alpha^{-1}\\
-(1-\rho_\circ) & \alpha^{-1}\leq \rho_\circ\leq 1.
\end{array}
\right.
\ee

\noindent
\fbox{$\rho_\ast\rightarrow 0$} We have for the current
\be\label{singul10}
J_\bullet(\rho_\bullet) =  \left\{
\begin{array}{ll}
\rho_\bullet(1-\rho_\bullet) & 0\leq \rho_\bullet\leq \beta, 1-\alpha^{-1}\leq \rho_\bullet \leq 1 \\
\beta(1-\alpha)+(\alpha-\beta)\rho_\bullet & \beta\leq \rho_\bullet\leq 1-\alpha.
\end{array}
\right.
\ee
\begin{figure}[h]
\begin{center}
	\begin{subfigure}[b]{0.4\linewidth}
\begin{tikzpicture}[scale=1.2]
\draw [thick,->] (-.25,0)--(3.5,0)node [below] {\scriptsize $\rho_\bullet$};
\draw [thick,->] (0,-.25)--(0,2)node [left] {\scriptsize $\frac{J_\bullet(\rho_\bullet)}{\beta}$};
\draw[dashed,domain=0:3,smooth,variable=\t]plot (\t,{\t*(3-\t)/2});
\def \n {0.9}
\draw[ultra thick,domain=0:\n,smooth,variable=\t]plot (\t,{\t*(3-\t)/2})--(3,0) node[below] {\scriptsize $1$};
\draw[dashed,fill=black] (\n,{\n*(3-\n)/2})--(\n,0)circle  (.03) node[below] {\scriptsize $\beta^{-1}$};
\end{tikzpicture}
\caption{}
\end{subfigure}
\hspace{1cm}
\begin{subfigure}[b]{0.4\linewidth}
\begin{tikzpicture}[scale=1.2]
\draw [thick,->] (-.25,0)--(3.5,0)node [below] {\scriptsize $\rho_\circ$};
\draw [thick,->] (0,-.25)--(0,2)node [left] {\scriptsize $-\frac{J_\circ(\rho_\circ)}{\alpha}$};
\def \r {1/2}
\def \n {2}
\draw[dashed,domain=0:3,smooth,variable=\t]plot (\t,{\t*(3-\t)*\r});
\draw[ultra thick,domain=0:{\n},smooth,variable=\t]plot (\t,{\t*(3-\t)*\r})--(3,0) node[below] {\scriptsize $1$};
\draw[dashed,fill=black] (\n,{\n*(3-\n)*\r})--(\n,0)circle  (.03) node[below] {\scriptsize $\alpha^{-1}$};
\end{tikzpicture}
\caption{}
\end{subfigure}

\vspace{.3cm}
\begin{subfigure}[b]{0.5\linewidth}
\begin{tikzpicture}[scale=1.2]
\draw [thick,->] (-.25,0)--(3.5,0)node [below] {\scriptsize $\rho_\bullet$};
\draw [thick,->] (0,-.25)--(0,2)node [left] {\scriptsize $J_\bullet(\rho_\bullet)$};
\def \r {1/2}
\def \n {.5}
\def \f {2}
\draw[dashed,domain=0:3,smooth,variable=\t]plot (\t,{\t*(3-\t)*\r})  node[below] {\scriptsize $1$};
\draw[ultra thick,domain=0:{\n},smooth,variable=\t]plot (\t,{\t*(3-\t)*\r})-- (\f,{\f*(3-\f)*\r});
\draw[ultra thick,domain={\f}:3,smooth,variable=\t]plot (\t,{\t*(3-\t)*\r});
\draw[dashed,fill=black] (\n,{\n*(3-\n)*\r})--(\n,0)circle  (.03) node[below] {\scriptsize $\beta$};
\draw[dashed,fill=black] (\f,{\f*(3-\f)*\r})--(\f,0)circle  (.03) node[below] {\scriptsize $1-\alpha$};
\end{tikzpicture}
\caption{}
\end{subfigure}
\end{center}
\caption{{\bf (a)} Current $J_\bullet$ for $\beta>1$ and $\rho_\circ=0$. {\bf (b)} Current $J_\circ$ for $\alpha>1$ and $\rho_\bullet=0$. 
{\bf (c)} Current $J_\bullet$ for $\alpha+\beta<1$ and $\rho_\ast=0$. 
The dashed lines correspond to the current for a strict absence of $\circ$--particles {\bf (a)}, $\bullet$--particles {\bf (b)}, $\ast$--particles {\bf (c)}.}
\end{figure}
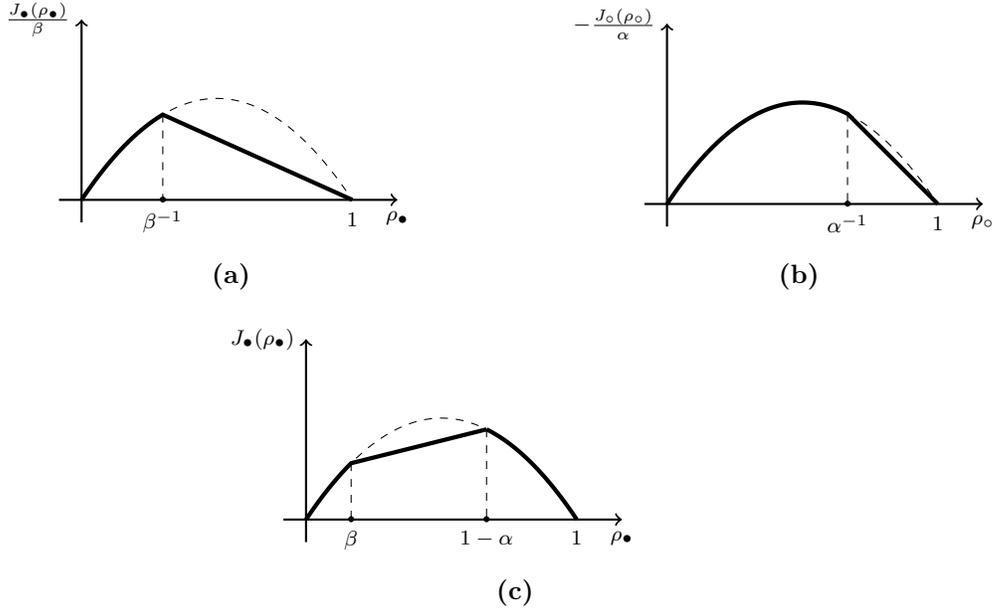
These results have to be compared to the situation in which we have \emph{strict} absence of a species of particles (and not just vanishing density).
Consider for example a system without $\circ$--particles. Such a system is effectively a single species TASEP with jump rates equal 
to $\beta$, with current just $J_\bullet(\rho_\bullet)=\beta \rho_\bullet(1-\rho_\bullet)$. Comparing this with eq.(\ref{singul1}) we see that for $\beta>1$ this behavior holds only for $0\leq \rho_\bullet\leq \beta^{-1}$, while for $\rho_\bullet > \beta^{-1}$
 the presence of even a single $0$--particle affects the macroscopic behavior of the system, giving rise to a different current.
 
At the boundary of $\mathcal{D}$, the average speed of the zero density particles displays also an interesting behavior.
\begin{itemize}[leftmargin=0cm] 
\item [] \fbox{$\rho_\circ\rightarrow 0$} Speed of $\circ$--particles 
\be\label{vel-singul1}
v_\circ(\rho_\bullet) =  \left\{
\begin{array}{ll}
-\frac{\alpha+\beta(1-\alpha)\rho_\bullet(1-\rho_\bullet)}{1+(\alpha-1)\rho_\bullet}-\beta \rho_\bullet & 0\leq \rho_\bullet\leq \beta^{-1}\\
-1 & \beta^{-1}\leq \rho_\bullet\leq 1,
\end{array}
\right.
\ee
\item [] \fbox{$\rho_\bullet \rightarrow 0$} Speed of $\bullet$--particles 
\be\label{vel-singul0}
v_\bullet(\rho_\circ) =  \left\{
\begin{array}{ll}
\frac{\beta+\alpha(1-\beta)\rho_\circ(1-\rho_\circ)}{1+(\beta-1)\rho_\circ}-\alpha \rho_\circ & 0\leq \rho_\circ\leq \alpha^{-1}\\
1 & \alpha^{-1}\leq \rho_\circ\leq 1,
\end{array}
\right.
\ee
\item [] \fbox{$\rho_\ast \rightarrow 0$} Speed of $\ast$--particles 
\be\label{vel-singul10}
v_\ast(\rho_\circ) =  z_\alpha-z_\beta
\ee
with
\be
z_\alpha = \left\{
\begin{array}{cc}
\alpha & \rho_\bullet\leq 1-\alpha\\ 
1-\rho_\bullet & \rho_\bullet \geq 1-\alpha
\end{array}
\right.,\qquad
z_\beta = \left\{
\begin{array}{cc}
\beta & \rho_\bullet\geq \beta\\ 
\rho_\bullet & \rho_\bullet \leq \beta
\end{array}
\right.
\ee
\end{itemize} 
The result in eqs.(\ref{vel-singul1}--\ref{vel-singul10}) have been obtained in the literature by considering systems with a single particle of either species: the cases $\rho_\ast\rightarrow 0$,  eqs.(\ref{singul10},\ref{vel-singul10}) first appeared  in  
\cite{mallick1996shocks,derrida1999bethe}, the cases $\rho_\circ \rightarrow 0$ or $\rho_\bullet \rightarrow  0$,  eqs.(\ref{singul1},\ref{vel-singul1}) and eqs.(\ref{singul0},\ref{vel-singul0}) the results first appeared in \cite{lee1997two}.

\section{Conservations laws}\label{sect:conserv}

Under Euler--scaling (where site position and time scale as $\epsilon^{-1}n,\epsilon^{-1}t$ for $\epsilon\rightarrow 0$) the density profiles are expected to evolve deterministically as solutions of a system of conservation laws.
Consider initial data $\rho_i^{(0)}(x)$ 
and to such data associate a family of initial conditions of  the 2-TASEP of product Bernoulli form, with local probability at site $n$ given by 
$$  
\mathbb{E}[\chi_i^{\epsilon}(n,t=0)] =  \rho_i^{(0)}(\epsilon n),  
$$
where $\chi_i^{\epsilon}(n,t)$ is the $i$--th species indicator function at time $t$ and site $n$.
We expect  that the random variable $\chi_i^{\epsilon}(\lfloor \epsilon^{-1}x \rfloor,\epsilon^{-1}t,)$ converges for $\epsilon \rightarrow 0$ to a deterministic density profile. More precisely we expect that
\be \label{converg}
\lim_{\epsilon \longrightarrow 0} \sum_{n:a\leq \epsilon n \leq b}\epsilon \,\chi_i^{\epsilon}(n,\epsilon^{-1}t) = \int_{a}^{b}\rho_i(x,t)dx, \qquad a.s.    
\ee
where   $\bfrho=(\rho_\circ,\rho_\bullet)$  is the solutions of a system of conservation lows 
\be \label{conserv-rho-j}
\partial_t \bfrho +\partial_x {\bf J}=0.
\ee
with initial condition $\bfrho(t=0)=\bfrho^{(0)}=(\rho^{(0)}_\circ,\rho^{(0)}_\bullet)$.
By making the usual hypothesis of \emph{local stationarity} we identify the local currents with the stationary currents at density ${\bfrho}$, given by eqs.(\ref{0Jrz},\ref{1Jrz}). 
 A more precise statement and proof of this result for the case $\alpha=\beta=\frac{1}{2}$ can be found in \cite{fritz2004derivation}. While the approach developed in \cite{fritz2004derivation} should extend to the full $\alpha+\beta=1$ line, for which the stationary measure is product, it is not clear to us whether that same approach could work for arbitrary values of $\alpha$ and $\beta$.
In the present paper we take  eqs.(\ref{converg},\ref{conserv-rho-j}) as working hypothesis. We work out the solutions of the system \eqref{conserv-rho-j} and compare them with Monte Carlo simulations. 

\subsection{The cases $\alpha=\beta=1$ and $\alpha+\beta=1$}

Before discussing the system of equation \eqref{conserv-rho-j} in full generality let us start with two particular cases.

\vspace{.3cm}
\noindent
\emph{ $\bullet\quad \alpha=\beta=1$}.

\noindent
When $\beta=1$, the $\bullet$--particles don't distinguish 
$\circ$--particles from $\ast$--particles. This means that the 
$\bullet$--particles evolve as in a single species TASEP. At the level of currents, for  $\beta=1$ we have indeed $J_\bullet=\rho_\bullet(1-\rho_\bullet)$. In this case the conservation law for $\rho_\bullet$ completely decouples from that of $\rho_\circ$ and takes the usual form of the non-viscous Burgers equation
$$
\partial_t \rho_\bullet +(1-2\rho_\bullet) \partial_x \rho_\bullet=0.
$$  
Analogously, for $\alpha =1$, the conservation law for $\rho_\circ$ completely decouples from that of $\rho_\bullet$ and takes the form
$$
\partial_t \rho_\circ -(1-2\rho_\circ) \partial_x \rho_\circ=0.
$$  
So for $\alpha=\beta=1$, system \eqref{conserv-rho-j} just decouples completely into two Burgers equations.

\vspace{.3cm}
\noindent
\emph{ $\bullet\quad \alpha+\beta=1$}.

\noindent
In this case, thanks to the factorization of the stationary measure,  the currents can be explicitly written as functions of the densities
$$
J_\circ=-\rho_\circ(\rho_\bullet+\beta(1-\rho_\circ-\rho_\bullet)),\qquad J_\bullet=\rho_\bullet(\rho_\circ+\alpha(1-\rho_\circ-\rho_\bullet)).
$$ 
One can consider conserved quantities $\rho$ and $v$, defined by 
\begin{equation}
\left(
\begin{array}{c}
\rho \\
v
\end{array}
\right) =
\left(\begin{array}{cc}
-\frac{\alpha(\alpha+2\beta)}{3}& -\frac{\beta(\alpha+2\beta)}{3}  \\
\alpha & -\beta
\end{array}
\right)\left(
\begin{array}{c}
\rho_\circ \\
\rho_\bullet
\end{array}
\right)+\left(
\begin{array}{c}
\frac{(2\alpha+\beta)(2\beta+\alpha)}{9} \\
\frac{\beta-\alpha}{3}
\end{array}
\right)
\end{equation}
The associated currents are (up to irrelevant additive constants)
$$
J_\rho= \rho v,\qquad J_v= \rho+v^2.
$$
These are the currents of the Leroux system \cite{leroux1978analyse,serre1988existence} (the particular case $\alpha=\beta=\frac{1}{2}$ is the one considered in \cite{fritz2004derivation}), which is known to be a Temple class system. 

\subsection{The general case: Riemann variables}

For the problem under investigation 
one could expect that on top of the generic difficulty of analyzing a coupled system of conservation equations, one has to face the additional complication due to the fact that the dependence of currents on the densities is not explicit but rather goes through the auxiliary variables $z_\alpha,z_\beta$.
%
%
Actually, quite unexpectedly the variables $\bfz$ turn out to be Riemann variables for our conservation laws, i.e. they diagonalize the system of eqs.(\ref{conserv-rho-j}) and simplify substantially their analysis.
From now on we want to think both $\bfrho$ and $\bfJ$ as functions of  
$\bfz$ .

We first notice that, solving eqs.(\ref{0Jrz},\ref{1Jrz}) for the densities $\bfrho$ in terms of 
$\bfz$ and $\bfJ$ and replacing them into eqs.(\ref{chang-var1},\ref{chang-var2}), the currents are the solution 
of the following linear system of equations
\begin{gather}\label{Jz0}
\frac{J_\circ}{z_\alpha}+\frac{J_\bullet}{z_\alpha-1}-\frac{J_\circ+J_\bullet}{z_\alpha-\alpha}+1=0\\\label{Jz1}
\frac{J_\bullet}{z_\beta}+\frac{J_\circ}{z_\beta-1}-\frac{J_\circ+J_\bullet}{z_\beta-\beta}+1=0.
\end{gather}
Now differentiate the l.h.s. of eq.(\ref{chang-var1}) with respect to $t$, differentiate the l.h.s. of eq.(\ref{Jz0}) with respect to $x$ and sum the obtained results.
%
Thanks to the conservation laws eqs.(\ref{conserv-rho-j}) the derivatives $\partial_t \bfrho$ and $\partial_x \bfJ$ cancel and 
one remains with
\begin{equation}\label{conserv-za}
\partial_t z_\alpha +v_\alpha(\bfz) \partial_x z_\alpha=0,\qquad v_\alpha(\bfz)=\frac{\left(\frac{J_\circ}{z_\alpha^2}+\frac{J_\bullet}{(z_\alpha-1)^2}-\frac{J_\circ+J_\bullet}{(z_\alpha-\alpha)^2}\right)}{\left(\frac{\rho_\circ}{z_\alpha^2}+\frac{\rho_\bullet}{(z_\alpha-1)^2}+\frac{1-\rho_\circ-\rho_\bullet}{(z_\alpha-\alpha)^2}\right)}.
\end{equation}
In the same way one obtains the equation for $z_\beta$
\begin{equation}\label{conserv-zb}
\partial_t z_\beta +v_\beta(\bfz) \partial_x z_\beta=0,\qquad v_\beta(\bfz)=\frac{\left(\frac{J_\bullet}{z_\beta^2}+\frac{J_\circ}{(z_\beta-1)^2}-\frac{J_\circ+J_\bullet}{(z_\beta-\beta)^2}\right)}{\left(\frac{\rho_\bullet}{z_\beta^2}+\frac{\rho_\circ}{(z_\beta-1)^2}+\frac{1-\rho_\circ-\rho_\bullet}{(z_\beta-\beta)^2}\right)}.
\end{equation}
The speeds $v_\alpha$ and $v_\beta$ are the eigenvalues of the linearization matrix $\partial_{\rho_j}J_i$, and on general grounds they can also be written as
\begin{equation}
v_\alpha= \partial_{\rho_i}J_i|_{z_\beta}=\frac{\partial_{z_\alpha}J_i(\bfz)}{\partial_{z_\alpha}\rho_i(\bfz)},\qquad v_\beta= \partial_{\rho_i}J_i|_{z_\alpha}=\frac{\partial_{z_\beta}J_i(\bfz)}{\partial_{z_\beta}\rho_i(\bfz)}.
\end{equation}
A close inspection of their expression allows to conclude that 
\begin{equation}\label{ineq-v}
v_\beta(\bfz)\geq v_\alpha(\bfz),
\end{equation}
with the equality holding for $1-z_\alpha-z_\beta=0$, which is a non empty set only for $\alpha+\beta\geq 1$. We conclude that for $\alpha+\beta<1$, the system in \eqref{conserv-rho-j} is strictly hyperbolic on the whole physical domain $\mathcal{D}$, whereas for for $\alpha+\beta\geq 1$ it is degenerate hyperbolic on the locus $1-z_\alpha-z_\beta=0$, i.e. $\rho_\ast=0,\rho_\circ\leq \alpha, \rho_\bullet\leq \beta$  (green segments in figures \ref{domains-maps-a},\ref{domains-maps-b}) and strictly hyperbolic on the rest of the physical domain. 

\begin{figure}[h!!]
\begin{center}
\begin{subfigure}[b]{0.3\linewidth}
\begin{tikzpicture}[scale=.4]
\draw[thick,->] (0,0) -- (7,0) node[below] {\scriptsize $\rho_\circ$};
\draw[thick,->] (0,0) -- (0,7) node[left] {\scriptsize $\rho_\bullet$};

\foreach \n in {0.2,0.4,0.6,0.8}
\foreach \x in {3}
\draw [thick, red, ->]  (0.3*\x, 6- 0.3*\x)-- (\x + 0.3*\x*\n - \x*\n, 6*\n- 0.3*\x*\n) ;

\foreach \x in {3}
\draw [thick, red] (\x,0) -- (0.3*\x, 6- 0.3*\x);

\foreach \n in {0.2,0.4,0.6,0.8}
	\foreach \y in {3}
	\draw [thick, blue ,->] (0,\y) -- (6*\n-0.4*\y*\n, \y + 0.4*\y*\n - \y*\n);
	\foreach \y in {3}
\draw [thick, blue] (0,\y) -- (6-0.4*\y,  0.4*\y );

\draw  (6*0.2,6-6*0.2) -- (6-0.4*6,0.4*6);
\draw (6*0.3,6-6*0.3) -- (0,6);
\draw (6,0) -- (6-0.4*6,0.4*6);

\draw [thick](2pt,6 cm) node[left] {\scriptsize $1$};
\draw [thick](6 cm,-2pt) node[below] {\scriptsize $1$};
\draw [thick, black!40!green ,->] (1.5,4.5) -- (2.3,6-2.3);
\draw [thick, black!40!green ,->] (2.3,6-2.3)--(3.1,6-3.1);
\draw [thick, black!40!green] (3.1,6-3.1)--(3.9,6-3.9);

\end{tikzpicture}
\caption{}
\end{subfigure}
\begin{subfigure}[b]{0.4\linewidth}
\begin{tikzpicture}[scale=3]


\fill [fill=green!40!white] (0.635714,1) -- (-0.25,1) --  (-0.25,0) node[below] {\scriptsize $v_\beta^-$} -- (0.635714,0) node[below] {\scriptsize $v_\beta^+$} ;

%

\fill [thick,fill=blue!40!white,variable=\y,domain=0.4:0,samples=100]

plot({(-0.04539 + 0.2856*\y - 0.537*\y^2 + 0.33*\y^3)/(-0.0714 + 0.204*\y - 
	0.165*\y^2)},{1 + (0.15*(-0.34 + 1.04*\y - 0.7*\y^2))/(0.102 - 0.165*\y)}) -- (0.635714,0) -- (-0.25,0) ;

\fill [thick, fill=yellow!40!white,variable=\y,domain=0.4:0,samples=100]

plot({(-0.04539 + 0.2856*\y - 0.537*\y^2 + 0.33*\y^3)/(-0.0714 + 0.204*\y - 
	0.165*\y^2)},{-((0.85*(0.21 - 0.3*\y)*\y)/(-0.102 + 0.165*\y))}) -- (0.635714,0) -- (-0.25,0)

(0.2,0.1) node {\scriptsize $\rho_\bullet$}
(0.4,0.4) node {\scriptsize $\rho_\ast$}
(0.4,0.8) node {\scriptsize $\rho_\circ$} ;

\draw[->] (0,0) -- (0,1.1) node[above] 
{\scriptsize { $\boldsymbol \rho$}};

\draw[->] (-1,0) -- (1,0) node[right] {\scriptsize $\xi$};
\end{tikzpicture}
\caption{}
\end{subfigure}
\end{center}
\caption{
(a) Projection in the $\boldsymbol \rho$--plane of the three types of rarefactions fans: in green a TASEP-like rarefaction fan, in blue an $\alpha$--fan, in red a $\beta$--fan. (b) A plot of a $\beta$--fan: $v_\beta^-=v_\beta(z_\alpha,1-z_\alpha), v_\beta^+=v_\beta(z_\alpha,0)$.
}\label{fig:fans}
\end{figure}
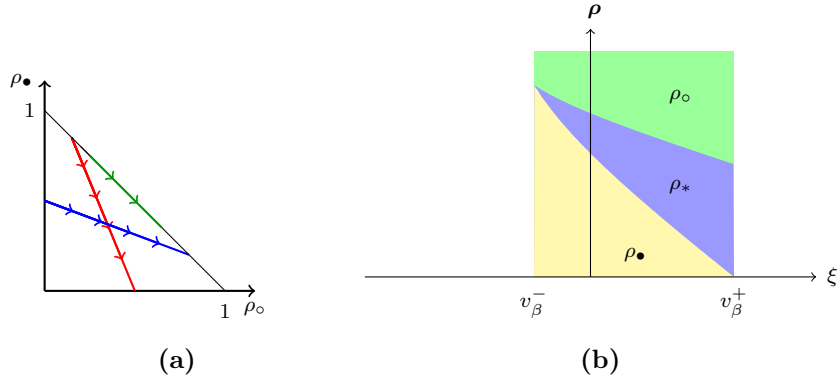

Using eqs.(\ref{conserv-za},\ref{conserv-zb}) we can easily work 
out  the \emph{continuous} solutions of eqs.(\ref{conserv-rho-j}), 
which depend only on the self-similarity variable $\xi=x/t$, 
$\bfz(x,t)=\bfz(\xi=x/t)$. They are solution of 
\begin{equation}
\begin{split}
(v_\alpha-\xi)\partial_\xi z_\alpha&=0\\
(v_\beta-\xi)\partial_\xi z_\beta&=0.
\end{split}
\end{equation}
Locally we have four possibilities. 
\begin{enumerate}
\item The trivial solution, namely both $z_\alpha$ and $z_\beta$ are constant.
\item Both $\partial_\xi z_\alpha\neq 0,\partial_\xi z_\beta\neq 0$.
In this case we must have  $v_\alpha-\xi=v_\beta-\xi=0$, and in particular $v_\alpha=v_\beta$. As mentioned above, this is possible only if $1-z_\alpha-z_\beta=0$. In this case we get a
\emph{rarefaction fan} of equation
\begin{equation}\label{single-fan}
\rho_\circ(\xi)=z_\alpha(\xi)=\frac{1+\xi}{2},\qquad \rho_\bullet(\xi)=z_\bullet(\xi)=\frac{1-\xi}{2}.
\end{equation}
Notice that the condition $1-z_\alpha-z_\beta=0$ implies absence of second class particles, and the solution \eqref{single-fan} corresponds to the fan solution of the single species TASEP.  
\item $z_\beta$ constant, $\partial_\xi z_\alpha\neq 0$. In this case we have a rarefaction fan ($\alpha$--fan) given in implicit form by 
$$
v_\alpha(z_\alpha(\xi,z_\beta),z_\beta)=\xi.
$$
One can show that at fixed $z_\beta$, $v_\alpha$ and $\rho_\circ$ are increasing function of $z_\alpha$. This means that $z_\alpha(\xi,z_\beta)$ and $\rho_\circ(\xi,z_\beta)$ are increasing functions of $\xi$.
\item $z_\alpha$ constant, $\partial_\xi z_\beta\neq 0$. In this case the rarefaction fan ($\beta$--fan) is given by 
$$
v_\beta(z_\alpha,z_\beta(\xi,z_\alpha))=\xi.
$$
At fixed $z_\alpha$,  $v_\beta$ is a decreasing function of $z_\beta$, while $\rho_\bullet$ is an increasing function of $z_\beta$ . This means that $z_\beta(\xi,z_\alpha)$ and $\rho_\bullet(\xi,z_\alpha)$ are decreasing functions of $\xi$.
\end{enumerate}
Projection in the $\boldsymbol \rho$--plane of the three types of fans as well as an example of a $\beta$--fan are represented in Figure \ref{fig:fans}.

\subsection{Shocks}

It is well known that a smooth solution of a nonlinear conservation 
laws like eqs.(\ref{conserv-rho-j}) may develop a shock discontinuity in finite time. One needs therefore to admit the notion of \emph{weak solution}, i.e. solution in the sense of distributions, which need not even be continuous. 
The discontinuity line is associate to a shock with trajectory $x_s(t)$, which has to satisfy the \emph{Rankine-Hugoniot jump relations }. If we denote by $[{\bfrho}]$ and $[{\bf J}]$ respectively the discontinuities of the densities and of the currents across the shock, i.e.
$$
[{\bfrho}]=\bfrho(x_s^+)-\bfrho(x_s^-),\qquad [{\bf J}]={\bf J}(x_s^+)-{\bf J}(x_s^-)
$$
then the Rankine-Hugoniot jump relations read 
\be\label{rankine-hugoniot}
[{\bfrho} ] -v_s [{\bf J}]=0,\qquad v_s=\text{shock's speed}.
\ee
The Rankine-Hugoniot jump relations, allow to express the speed of the shock in terms of the discontinuity and at the same time put a constraint on the admissible discontinuities, the Hugoniot condition:
\be\label{hugoniot-condition}
\det 
\left(
\begin{array}{cc}
[\rho_\circ ] &  [J_\circ]\\[5pt]
[\rho_\bullet]  & [J_\bullet]
\end{array}
\right)=0.
\ee
In order to analyze the Hugoniot condition, we consider  $\bfrho(x_s^-)=\bfrho^-$ as fixed.
For strictly hyperbolic systems of $N$ conservation laws, it is known that the set of $\bfrho^+=\bfrho(x_s^+)$  satisfying the Hugoniot condition passes through $\bfrho^-$ and locally decompose around $\bfrho^-$  in $N$ different branches \cite{lefloch2002hyperbolic}, each one called a \emph{shock curve}. In our case we expect then two shock curves passing through any point $\rho^-$, which correspond to two different kinds of shocks.
Using eq.(\ref{chang-var1}) and eq.(\ref{Jz0}) we see that if $z_\alpha^{+}=z_\alpha^{-}=z_\alpha$, then we have 
\begin{equation*}
\begin{split}
\frac{[\rho_\circ]}{z_\alpha}+\frac{[\rho_\bullet]}{z_\alpha-1}-\frac{[\rho_\circ]+[\rho_\bullet]}{z_\alpha-\alpha}=0\\
\frac{[J_\circ]}{z_\alpha}+\frac{[J_\bullet]}{z_\alpha-1}-\frac{[J_\circ]+[J_\bullet]}{z_\alpha-\alpha}=0.
\end{split}
\end{equation*}
This means that the matrix in the  eq.\eqref{hugoniot-condition} 
 has the left null vector $(\frac{1}{z_\alpha}-\frac{1}{z_\alpha-\alpha},\frac{1}{z_\alpha-1}-\frac{1}{z_\alpha-\alpha}) $ and therefore its determinant vanishes. We conclude that a straight line
  at $z_\alpha$ constant is a shock curves. In the same way we find 
  that the lines at $z_\beta$ constant are also shock curves.
  In conclusion we have found that we have two kind of admissible shocks
\begin{itemize}
\item $\beta$-shocks: $z_\alpha(\bfrho^-)=z_\alpha(\bfrho^+)=z_\alpha$ with speed $v_{s,\beta}(z_\alpha;z_\beta^-,z_\beta^+)$;
\item $\alpha$-shocks: $z_\beta(\bfrho^-)=z_\beta(\bfrho^+)=z_\beta$, with speed $v_{s,\alpha}(z_\beta;z_\alpha^-,z_\alpha^+)$.
\end{itemize}
The corresponding shocks speeds $v_{s,\beta}(z_\alpha;z_\beta^-,z_\beta^+)$ and $v_{s,\alpha}(z_\beta;z_\alpha^-,z_\alpha^+)$ do not have particularly transparent expressions except for some particular cases. 
In the case $\alpha=\beta=1$, a $\beta$--shock is a discontinuity of the density $\rho_\bullet$, with $\rho_\circ$ constant across the discontinuity, while an $\alpha$--shock is the other way round, i.e. a discontinuity of the density $\rho_\circ$, with $\rho_\bullet$ constant across the discontinuity.
Shock curves coincide with rarefaction curves so this an example of a conservation system of Temple class \cite{temple1983systems}.

A further analysis of the Hugoniot condition allows to conclude that in the bulk of the physical domain $\mathcal{D}$, the only possible shocks are $\alpha$ and $\beta$--shocks.  
There exists however one more class of shocks when both sides of the discontinuity lie on the boundary line $\rho_\ast=0$. These shocks have speed  
$v_s=\frac{[J_\bullet]}{[\rho_\bullet]}$,
where the current $J_\bullet$ is given by eq.\eqref{singul10}.
%
%
%
%
%
%
%
%
%
%
%
%
%

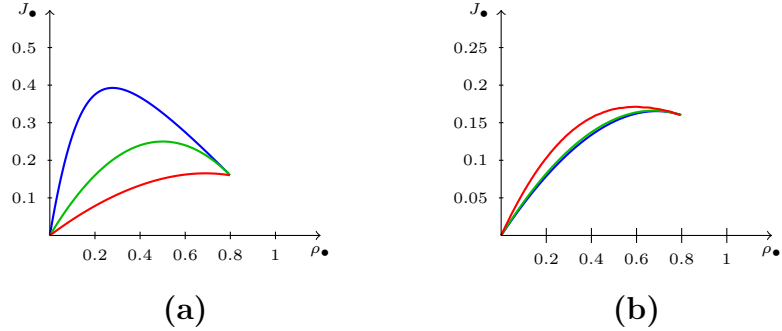
\begin{figure}[h!!]
\begin{center}
\begin{tikzpicture}[xscale=3,yscale=5]
\def\a{4};
\def\b{5};
\def\za{.2};
\draw [blue, thick] plot[smooth,domain=0:{1-\za}, samples=80] ({-( (-1 + \za)*\x*(-\a + \za - \b*\za + \a*\x))/(-\a*\b + \a*\b*\za + \a*\b*\x + \za*\x - \a*\za*\x - \b*\za*\x)},
{
(
(-1 +\za)*\x*
(\a*\b + \a*\za - \a*\b*\za - (\za)^2 + \b*(\za)^2 - 
\a*\x - \a*\b*\x + \a*(\x)^2)
)/(-\a*\b 
+ \a*\b*\za + \a*\b*\x + \za*\x - \a*\za*\x - \b*\za*\x)});
\def\a{4};
\def\b{1};
\def\za{.2};
\draw [black!25!green, thick] plot[smooth,domain=0:{1-\za}, samples=80] ({-( (-1 + \za)*\x*(-\a + \za - \b*\za + \a*\x))/(-\a*\b + \a*\b*\za + \a*\b*\x + \za*\x - \a*\za*\x - \b*\za*\x)},
{
(
(-1 +\za)*\x*
(\a*\b + \a*\za - \a*\b*\za - (\za)^2 + \b*(\za)^2 - 
\a*\x - \a*\b*\x + \a*(\x)^2)
)/(-\a*\b 
+ \a*\b*\za + \a*\b*\x + \za*\x - \a*\za*\x - \b*\za*\x)});
\def\a{4};
\def\b{.3};
\def\za{.2};
\draw [red, thick] plot[smooth,domain=0:{\b}, samples=80] ({-( (-1 + \za)*\x*(-\a + \za - \b*\za + \a*\x))/(-\a*\b + \a*\b*\za + \a*\b*\x + \za*\x - \a*\za*\x - \b*\za*\x)},
{
(
(-1 +\za)*\x*
(\a*\b + \a*\za - \a*\b*\za - (\za)^2 + \b*(\za)^2 - 
\a*\x - \a*\b*\x + \a*(\x)^2)
)/(-\a*\b 
+ \a*\b*\za + \a*\b*\x + \za*\x - \a*\za*\x - \b*\za*\x)});
\draw [->] (0,0)-- (1.2,0) node[below] {\tiny $\rho_\bullet$};
\draw [->] (0,0)-- (0,.6) node [left] {\tiny $J_\bullet$};

\foreach \y in {0.2,0.4,0.6,0.8,1} \draw ({\y},-.01) node[below]{\tiny \y}--({\y},+.01);
\foreach \y in {0.1,0.2,0.3,0.4,0.5} \draw (-.01,{\y}) node[left]{\tiny \y}--(+.01,{\y});
\draw (0.6,-.2) node {\bf(a)};
\begin{scope}[xshift = 2cm,yscale=2]
\def\a{3};
\def\b{.3};
\def\za{.2};
\draw [blue, thick] plot[smooth,domain=0:{\b}, samples=80] ({-( (-1 + \za)*\x*(-\a + \za - \b*\za + \a*\x))/(-\a*\b + \a*\b*\za + \a*\b*\x + \za*\x - \a*\za*\x - \b*\za*\x)},
{
(
(-1 +\za)*\x*
(\a*\b + \a*\za - \a*\b*\za - (\za)^2 + \b*(\za)^2 - 
\a*\x - \a*\b*\x + \a*(\x)^2)
)/(-\a*\b 
+ \a*\b*\za + \a*\b*\x + \za*\x - \a*\za*\x - \b*\za*\x)});
\def\a{1};
\def\b{.3};
\def\za{.2};
\draw [black!25!green, thick] plot[smooth,domain=0:{\b}, samples=80] ({-( (-1 + \za)*\x*(-\a + \za - \b*\za + \a*\x))/(-\a*\b + \a*\b*\za + \a*\b*\x + \za*\x - \a*\za*\x - \b*\za*\x)},
{
(
(-1 +\za)*\x*
(\a*\b + \a*\za - \a*\b*\za - (\za)^2 + \b*(\za)^2 - 
\a*\x - \a*\b*\x + \a*(\x)^2)
)/(-\a*\b 
+ \a*\b*\za + \a*\b*\x + \za*\x - \a*\za*\x - \b*\za*\x)});
\def\a{.3};
\def\b{.3};
\def\za{.2};
\draw [red, thick] plot[smooth,domain=0:{\b}, samples=80] ({-( (-1 + \za)*\x*(-\a + \za - \b*\za + \a*\x))/(-\a*\b + \a*\b*\za + \a*\b*\x + \za*\x - \a*\za*\x - \b*\za*\x)},
{
(
(-1 +\za)*\x*
(\a*\b + \a*\za - \a*\b*\za - (\za)^2 + \b*(\za)^2 - 
\a*\x - \a*\b*\x + \a*(\x)^2)
)/(-\a*\b 
+ \a*\b*\za + \a*\b*\x + \za*\x - \a*\za*\x - \b*\za*\x)});
\draw [->] (0,0)-- (1.2,0) node[below] {\tiny $\rho_\bullet$};
\draw [->] (0,0)-- (0,.3) node [left] {\tiny $J_\bullet$};

\foreach \y in {0.2,0.4,0.6,0.8,1} \draw ({\y},-.01) node[below]{\tiny \y}--({\y},+.01);
\foreach \y in {0.05,0.1,0.15,0.2,0.25} \draw (-.01,{\y}) node[left]{\tiny \y}--(+.01,{\y});
\draw (0.6,-.1) node {\bf (b)};
\end{scope}
\end{tikzpicture}
\end{center}
\caption{Current $J_\bullet$ as function of $\rho_\bullet$ at constant $z_\alpha=.2$. (a) Fixed $\alpha=4$ and decreasing $\beta$ (blue line $\beta=5$, green line $\beta=1$, red line $\beta=0.3$). (b) Fixed $\beta=0.3$ and decreasing $\alpha$ (blue line $\alpha=3$, green line $\alpha=1$, red line $\alpha=0.3$).  
}\label{fig:current-za-const}
\end{figure}

It is possible to show that at fixed $z_\alpha$ the current $J_\bullet$ is a concave function of the density $\rho_\bullet$ (see fig. \ref{fig:current-za-const} ). 
This implies that, for a fixed value of the
densities on one side of the shock, say $\bfrho^-$, the speed of a $\beta$--shock is a decreasing function of $\rho_\bullet^+$.
This property can be conveniently reformulated in terms of the $\bfz$ variables. 
At constant $z_\alpha$, $\rho_\bullet$ is an increasing function of $z_\beta$, hence at fixed $z_\alpha$ and $z_\beta^-$, 
the speed of $\beta$--shock $v_{s,\beta}(z_\alpha;z_\beta^-,z^+_\beta)$ is a decreasing function of $z^+_\beta$. In particular it takes its minimum for the largest $z_\beta$ allowed, i.e. 
 $z_{\beta,\textrm{max}}=1-z_\alpha$
\begin{equation}
v_{s,\beta}(z_\alpha;z_\beta,z^+_\beta)\geq v_{s,\beta}(z_\alpha;z_\beta,1-z_\alpha) \qquad 0\leq z^+_\beta \leq 1-z_\alpha.
\end{equation}
The current $J_\circ$ is a convex function of $\rho_\circ$ at fixed $z_\beta$. A similar reasoning as the one presented above allows to  
conclude 
\begin{equation}
v_{s,\alpha}(z_\beta;z_\alpha,z^+_\alpha)\leq v_{s,\alpha}(z_\beta;z_\alpha,1-z_\beta)\qquad 0\leq  z^+_\alpha\leq  1-z_\beta.
\end{equation}
Now, from an explicit computation, we notice that 
$$
v_{s,\alpha}(z_\beta;z_\alpha,1-z_\beta)=v_{s,\beta}(z_\alpha;z_\beta,1-z_\alpha)=z_\alpha-z_\beta.
$$
This allows to conclude that 
\be\label{ineq-shock-speeds}
v_{s,\alpha}(z_\beta;z_\alpha,z^*_\alpha)\leq  v_{s,\beta}(z_\alpha;z_\beta,z^*_\beta).
\ee
In words this means that, for a fixed value of the densities on one side of the shock, the speed of any $\beta$--shock is larger than the speed of any $\alpha$--shock.

\noindent
Since $\lim_{z_\alpha^*\rightarrow z_\alpha} v_{s,\alpha}(z_\beta;z_\alpha,z^*_\alpha)=v_{\alpha}(z_\alpha,z_\beta)$ and 
$\lim_{z_\beta^*\rightarrow z_\beta}v_{s,\beta}(z_\alpha;z_\beta,z^*_\beta)=v_{\beta}(z_\alpha,z_\beta)$, we get also
\begin{gather}\label{ineq-v-vshA}
v_{\alpha}(z_\alpha,z_\beta)\leq  v_{s,\beta}(z_\alpha;z_\beta,z^*_\beta) \\\label{ineq-v-vshB}
v_{\beta}(z_\alpha,z_\beta)\geq v_{s,\alpha}(z_\beta;z_\alpha,z^*_\alpha).
\end{gather}

The Hugoniot condition is not sufficient to select the physical shocks. Indeed eqs.\eqref{conserv-rho-j} has to be thought of as the zero viscosity limit of a set of conservation laws which contains a diffusive term, a term which comes from the  microscopic corrections to the currents and depends on the derivatives of the densities.
Inviscid limits of viscous solutions are typically characterized by entropy conditions, which for shocks take the form of the Liu entropy criterion \cite{lefloch2002hyperbolic}. Let the right densities $\bfrho^+$ lie on a Hugoniot curve emanating from $\bfrho^-$, then for all densities  
$\bfrho^* $ lying on the same Hugoniot curve in between $\bfrho^-$ and $\bfrho^+$ the Liu condition states that 
\be\label{liu-condition-general}
v_s(\bfrho^-,\bfrho^+)\leq v_s(\bfrho^-,\bfrho^*).
\ee
Physically this condition can be understood as a stability condition: if under a perturbation the shock were to split by inserting an intermediate state $\bfrho^*$, then a violation of  condition  \eqref{liu-condition-general} would imply that the shock between $\bfrho^-$ and $\bfrho^*$  would move away from the original shock between $\bfrho^-$ and $\bfrho^+$.
In the case of $\beta$--shocks the Liu condition reads
\be\label{liu-cond-b}
v_{s,\beta}(z_\alpha;z_\beta^-,z_\beta^+)  \leq v_{s,\beta}(z_\alpha;z_\beta^-,z_\beta^*),\qquad \forall z_\beta \in [\min(z_\beta^-,z_\beta^+),\max(z_\beta^-,z_\beta^+)]
\ee
Since, as mentioned above, at fixed $z_\alpha$ the current $J_\beta$ is a concave function of the density $\rho_\bullet$, $
\partial_{\rho_\bullet}^2 J_\beta|_{z_\alpha} <0
$, we conclude that the Liu constraint means 
$\rho_\bullet^-<\rho_\bullet^+$. This can also formulated in terms of the $z$ variables. Indeed, since $\partial_{z_\beta} \rho_\bullet|_{z_\alpha}>0$, we must have  $z_\beta^-<z_\beta^+$.
A similar analysis can be performed for $\alpha$--shocks. 

\vspace*{.2cm}
\noindent
\begin{minipage}{0.7 \textwidth}
Here is a summary of our results. They are schematized in the figure on the right, where  we have reported the direction of the possible shock--discontinuities.
\begin{itemize}
\item For $\alpha$--shocks (blue oriented line), we need $\rho_\circ^->\rho_\circ^+$ or equivalently $z_\alpha^{-}>z_\alpha^{+}$. 
\item For $\beta$--shocks (red oriented line), we need $\rho_\bullet^-<\rho_\bullet^+$ or equivalently $z_\beta^{-}<z_\beta^{+}$. 
\end{itemize}
\end{minipage}
 \begin{minipage}{0.3 \textwidth}
 
\begin{center}
 \begin{tikzpicture}[scale=0.4]
\draw[thick,->] (0,0) -- (7,0) node[below] {\scriptsize $\rho_\circ$};
\draw[thick,->] (0,0) -- (0,7) node[left] {\scriptsize  $\rho_\bullet$};

\foreach \n in {0.2,0.4,0.6,0.8}
\foreach \x in {3}
\foreach \x in {3}
\draw [thick, red, ->, dashed] (\x,0) -- (\x + 0.3*\x*\n - \x*\n,  6*\n- 0.3*\x*\n );

\foreach \x in {3}
\draw [thick, red, dashed] (\x,0) -- (0.3*\x, 6- 0.3*\x);

\foreach \n in {0.2,0.4,0.6,0.8}
\foreach \y in {3}
\draw [thick, blue ,->, dashed] (6-0.4*\y, 0.4*\y ) -- (6*\n-0.4*\y*\n, \y + 0.4*\y*\n - \y*\n);

\foreach \y in {3}
\draw [thick, blue, dashed] (0,\y) -- (6-0.4*\y,  0.4*\y );

\draw  (6*0.2,6-6*0.2) -- (6-0.4*6,0.4*6);
\draw (6*0.3,6-6*0.3) -- (0,6);
\draw (6,0) -- (6-0.4*6,0.4*6);

\draw [thick](2pt,6 cm) node[left] {\scriptsize $1$};
\draw [thick](6 cm,-2pt) node[below] {\scriptsize $1$};

\end{tikzpicture}
\end{center}

\end{minipage}

\noindent

\subsection{Riemann's problem}\label{sect-riemann}

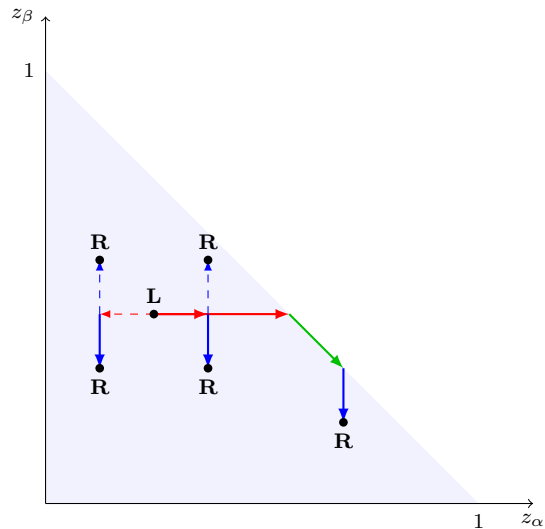
\begin{figure}[h]
\begin{center}
\begin{tikzpicture}[scale= 1.2]
\begin{scope}[scale = .6, xshift=10cm]
	\fill [fill=white!95!blue] (-1,-1)--(-1,7) node [left] {\scriptsize $1$}--(7,-1)node [below] {\scriptsize $1$}--cycle;
	\draw[->] (-1,-1) -- (8,-1) node[below] {\scriptsize $z_{\alpha}$};
	\draw[->] (-1,-1) -- (-1,8) node[left] {\scriptsize  $z_{\beta}$};
	\begin{scope}[shift={(-1,.5)}]
	\draw [->,>=latex,dashed,red](2,2) --(1,2);
	\draw [->,>=latex,dashed,blue](1,2)--(1,3);
	\draw [->,>=latex,thick,blue](1,2)--(1,1);
	\draw [->,>=latex,thick,red](2,2) --(3,2);
	\draw [->,>=latex,thick,red](2,2) --(4.5,2);
	\draw [->,>=latex,thick,black!25!green](4.5,2) --(5.5,1);
	\draw [->,>=latex,thick,blue](5.5,1)--(5.5,0);
	\draw [->,>=latex,thick,blue](3,2)--(3,1);
	\draw [->,>=latex,dashed,blue](3,2)--(3,3);
	\fill (2,2) circle (.08) node[above] {\scriptsize \bf L};
	\fill (1,3) circle (.08) node[above] {\scriptsize \bf R};
	\fill (1,1) circle (.08) node[below] {\scriptsize \bf R};
	\fill (3,1) circle (.08) node[below] {\scriptsize \bf R};
	\fill (5.5,0) circle (.08) node[below] {\scriptsize \bf R};
	\fill (3,3) circle (.08) node[above] {\scriptsize \bf R};
	\end{scope}
	
	\end{scope}
\end{tikzpicture}
\end{center}
\caption{Projection in the $\bfz$--plane of the different type of solutions of the Riemann problem. The left values of the $\bfz$  variables are represented by the point $L$. The points $R$ represent the different distinct possibilities for the right values of the $\bfz$  variables. Continuous lines represent fans, while dashed lines represent shock. In green is indicated the possible TASEP-like fan, in red either an $\alpha$--shock or an $\alpha$--fan and in blue either a $\beta$--shock or a $\beta$--fan. 
Starting from the values in L}\label{fig:riemann}
\end{figure}

With the result for the rarefaction curves and the shock curves at our disposal, it is rather simple to describe the general solution of the Riemann problem, i.e. the
solution of eqs.\eqref{conserv-rho-j}  with domain wall initial conditions
\be
\bfrho(x,t=0)=\left\{
\begin{array}{l}
\bfrho^{L}=(\rho_\circ^L,\rho_\bullet^L)\qquad x<0\\
\bfrho^R=(\rho_\circ^R,\rho_\bullet^R)\qquad x>0.
\end{array}
\right.
\ee
By uniqueness, the solution of the Riemann problem has to take the form $\bfrho(x,t)=\bfrho(\xi)$ with $\xi=x/t$ and is given by a sequence of rarefaction waves and/or shocks.
It is best described in terms of the variables $\bfz^L=(z_\alpha^L,z_\beta^L), \bfz^R=(z_\alpha^R,z_\beta^R)$.
We have four possible situations (these are schematically summarized in figure \ref{fig:riemann}).

\vspace{.3cm}
\noindent
$\bullet$ $z^L_\alpha>z^R_\alpha, z^L_\beta<z^R_\beta$. The solution is composed of two shocks: an $\alpha$-shock with $\bfz^-=(z^L_\alpha,z^L_\beta)$ and $\bfz^+=(z^R_\alpha,z^L_\beta)$  at position $\xi_\alpha= v_{s,\alpha}(z^L_\beta;z^L_\alpha,z^R_\alpha)$, 
followed by a $\beta$--shock with $\bfz^-=(z^R_\alpha,z^L_\beta)$ and $\bfz^+=(z^R_\alpha, z^R_\beta)$ at position $\xi_\beta=v_{s,\beta}(z^R_\alpha;z^L_\beta,z^R_\beta)$. 
This result follows from the inequality \eqref{ineq-shock-speeds}
$$
\xi_\alpha= v_{s,\alpha}(z^L_\beta;z^L_\alpha,z^R_\alpha)\leq v_{s,\beta}(z^R_\alpha;z^L_\beta,z^R_\beta) =\xi_\beta.
$$

\vspace{.3cm}
\noindent
$\bullet$ $z^L_\alpha>z^R_\alpha, z^L_\beta>z^R_\beta$. The solution is composed of an $\alpha$-shock and a $\beta$--fan.
The $\alpha$--shock has $\bfz^-=(z^L_\alpha,z^L_\beta)$ and $\bfz^+=(z^R_\alpha,z^L_\beta)$    and it is located at position $\xi_\alpha= v_{s,\alpha}(z^L_\beta;z^L_\alpha,z^R_\alpha)$.
 The $\beta$--fan starts with value $(z^R_\alpha,z^L_\beta)$ at $\xi_{\beta,1}= v_{\beta}(z^R_\alpha,z^L_\beta)$ and ends  with value $(z^R_\alpha,z^R_\beta)$ at $\xi_{\beta,2}= v_{\beta}(z^R_\alpha,z^R_\beta)$. 
This result follows from the inequality 
$$
\xi_\alpha= v_{s,\alpha}(z^L_\beta;z^L_\alpha,z^R_\alpha)\leq v_{\beta}(z^R_\alpha,z^L_\beta) =\xi_{\beta,1} \leq v_{\beta}(z^R_\alpha,z^R_\beta)=\xi_{\beta,2}.
$$
The first inequality is just inequality \eqref{ineq-v-vshB}, while the second one follows from the fact that  $v_{\beta}$ is a decreasing function of $z_\beta$.

\vspace{.2cm}
\noindent
$\bullet$ $z^L_\alpha<z^R_\alpha, z^L_\beta<z^R_\beta$. The solution is composed of an $\alpha$--fan and a $\beta$--shock.
The $\alpha$--fan starts with value $(z^L_\alpha,z^L_\beta)$ at $\xi_{\alpha,1}=v_{\alpha}(z^L_\alpha,z^L_\beta)$ and ends with value $(z^R_\alpha,z^L_\beta)$ at $\xi_{\alpha,2}=v_{\alpha}(z^R_\alpha,z^L_\beta)$. 
The $\beta$ shock has $\bfz^-=(z^R_\alpha,z^L_\beta)$ and $\bfz^+=(z^R_\alpha, z^R_\beta)$ and is located at position $\xi_\beta=v_{s,\beta}(z^R_\alpha;z^L_\beta,z^R_\beta)$.

\vspace{.2cm}
\noindent
$\bullet$ $z^L_\alpha<z^R_\alpha, z^L_\beta>z^R_\beta$. The solution is composed of fans. One has to distinguish two cases depending whether $z_\alpha^R+z_\beta^L$ is larger or smaller than $1$. 
If $z_\alpha^R+z_\beta^L<1$, the solution consists of an $\alpha$--fan starting with value $(z_\alpha^L,z_\beta^L)$ at $\xi_{\alpha,1}=v_{\alpha}(z^L_\alpha,z^L_\beta)$
and ending with value $(z_\alpha^R,z_\beta^L)$ at $\xi_{\alpha,2}=v_{\alpha}(z^R_\alpha,z^L_\beta)$ followed by  a $\beta$--fan starting at $\xi_{\beta,1}=v_{\beta}(z^R_\alpha,z^L_\beta)$ and ending at $\xi_{\beta,1}=v_{\beta}(z^R_\alpha,z^R_\beta)$.
If $\alpha+\beta>1$ then it is possible to have $z_\alpha^R+z_\beta^L>1$. In this case the $\alpha$-fan cannot reach the value $(z_\alpha^R,z_\beta^L)$, which lies outside the physical domain. It ends at the value $(1-z_\beta^L,z_\beta^L)$, followed by a degenerate fan till $(z_\alpha^R,1-z_\alpha^R)$ and then by a $\beta$--fan till  $(z_\alpha^R,z_\beta^R)$.   
%
%

\section{Monte Carlo simulations}\label{sec:montecarlo}

We come back to the original stochastic model and compare the predictions coming from the hydrodynamic equations with numerical simulations. 
We have simulated our model on a finite lattice of half-integer coordinates $[-L,L]$ ($L=2100$). The system is initialized in a random configuration sampled from a product Bernoulli measure corresponding to local densities $\bfrho^{L}$ on negative sites and $\bfrho^{R}$.
This means that the hydrodynamic prediction is expected to be meaningful until the kinetic wave coming from the origin meets those coming from the from the boundary meets. In this sense  the specific choice of boundary dynamics is irrelevant.

%
From our assumption eq.(\ref{converg}), it follows that for large time the integrated densities
%
%
%
\begin{equation}
h_{i}\left(\frac{n}{t},t\right) := \frac{1}{t} \sum_{-t < k \leq n} \chi_i(k,t)\quad \quad  - t < n < t
\end{equation}
should converge to the deterministic functions
%
\begin{equation}
h_{i}(u) = \int_{-1}^{u} \rho_{i}(\xi) d \xi,
\end{equation}
where  $\rho_{i}(\xi)$ are the solutions of the Riemann problem found in Section \ref{sect-riemann}.
In what follows we show the results of some of our simulations.
On the left pictures we report the plots of simulated the heights functions (dashed lines) together with the predicted ones (continuous lines). For illustration we have also plotted the predicted densities and on the right their projections on the $\bfrho$--plane. 
We  clearly see a confirmation of the predictions coming from the solution of the hydrodynamic equations.

\vspace{.3cm}
\noindent

\begin{figure}[H]
	\begin{subfigure}{.55\textwidth}
		\centering
		\begin{tikzpicture}
		\node (0,0) {\includegraphics[width=7cm]{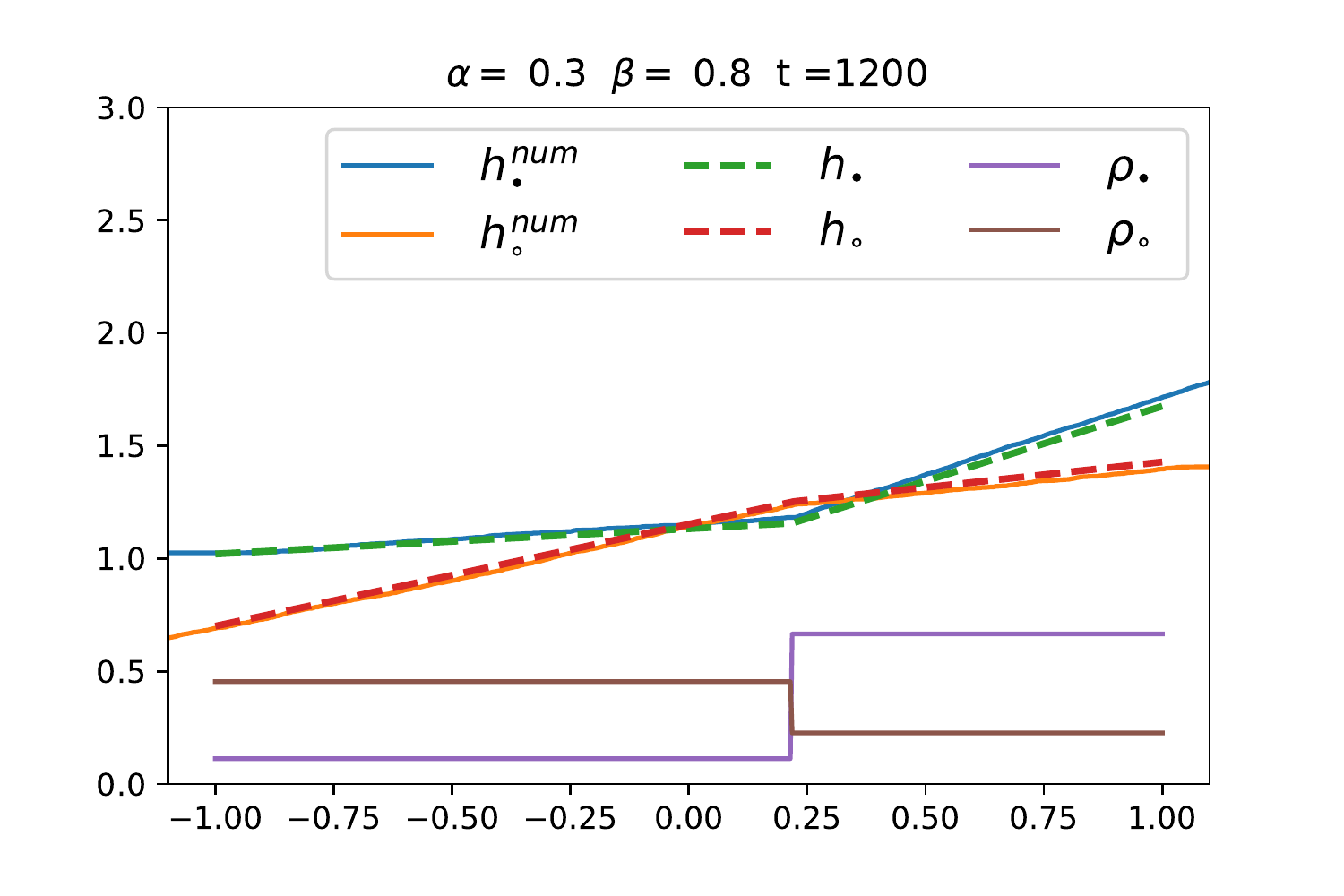}};
		\end{tikzpicture}
		\label{fig:sub3}
	\end{subfigure}
		\begin{subfigure}{.44\textwidth}
		\centering
		
		\begin{tikzpicture}
		
		\begin{scope}[scale = .5]
		
		\draw[thick,->] (0,0) -- (7,0) node[below] {\scriptsize $\rho_\circ$};
		\draw[thick,->] (0,0) -- (0,7) node[left] {\scriptsize $\rho_\bullet$};

		\draw [line width=0.7 mm, red, ->, dashed] (0.45*6,0.112*6) -- ( 0.45*3+0.226*3, 0.112*3+0.665*3) ;
		\draw [line width=0.7 mm, red, dashed]  ( 0.45*3+0.226*3, 0.112*3+0.665*3) -- (0.226*6,0.665*6);

		\node[circle,fill=black,inner sep=0pt,minimum size=5pt,label=west:{\scriptsize $L$} ]  at (0.45*6,0.112*6)  {};
		
		\node[circle,fill=black,inner sep=0pt,minimum size=5pt,label=west:{\scriptsize $R$} ]  at (0.226*6,0.665*6)  {};

		\draw  (6*0.2,6-6*0.2) -- (6-0.4*6,0.4*6);
		\draw (6*0.3,6-6*0.3) -- (0,6);
		\draw (6,0) -- (6-0.4*6,0.4*6);
		
		\draw [thick](2pt,6 cm) node[anchor=east] {\scriptsize $1$};
		\draw [thick](6 cm,-2pt) node[anchor=north] {\scriptsize $1$};
		
		\end{scope}
		
		\end{tikzpicture}

		\label{fig:sub1}
	\end{subfigure}%
\caption{Situation with a single shock}
\end{figure}
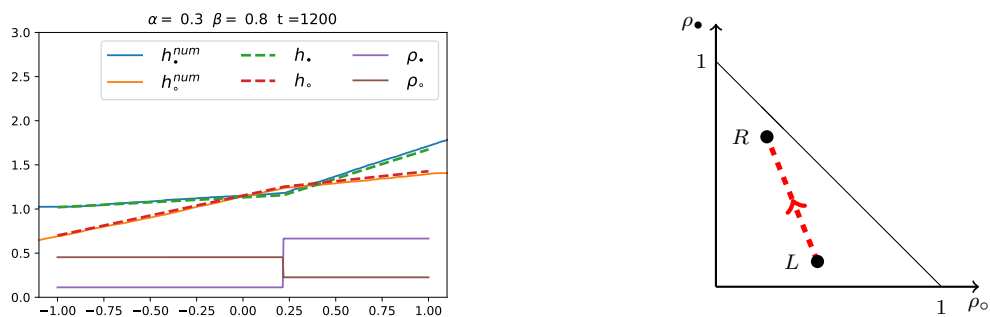
	
\begin{figure}[H]
	
		\begin{subfigure}{.55\textwidth}
		\centering
		
		\includegraphics[width=7cm]{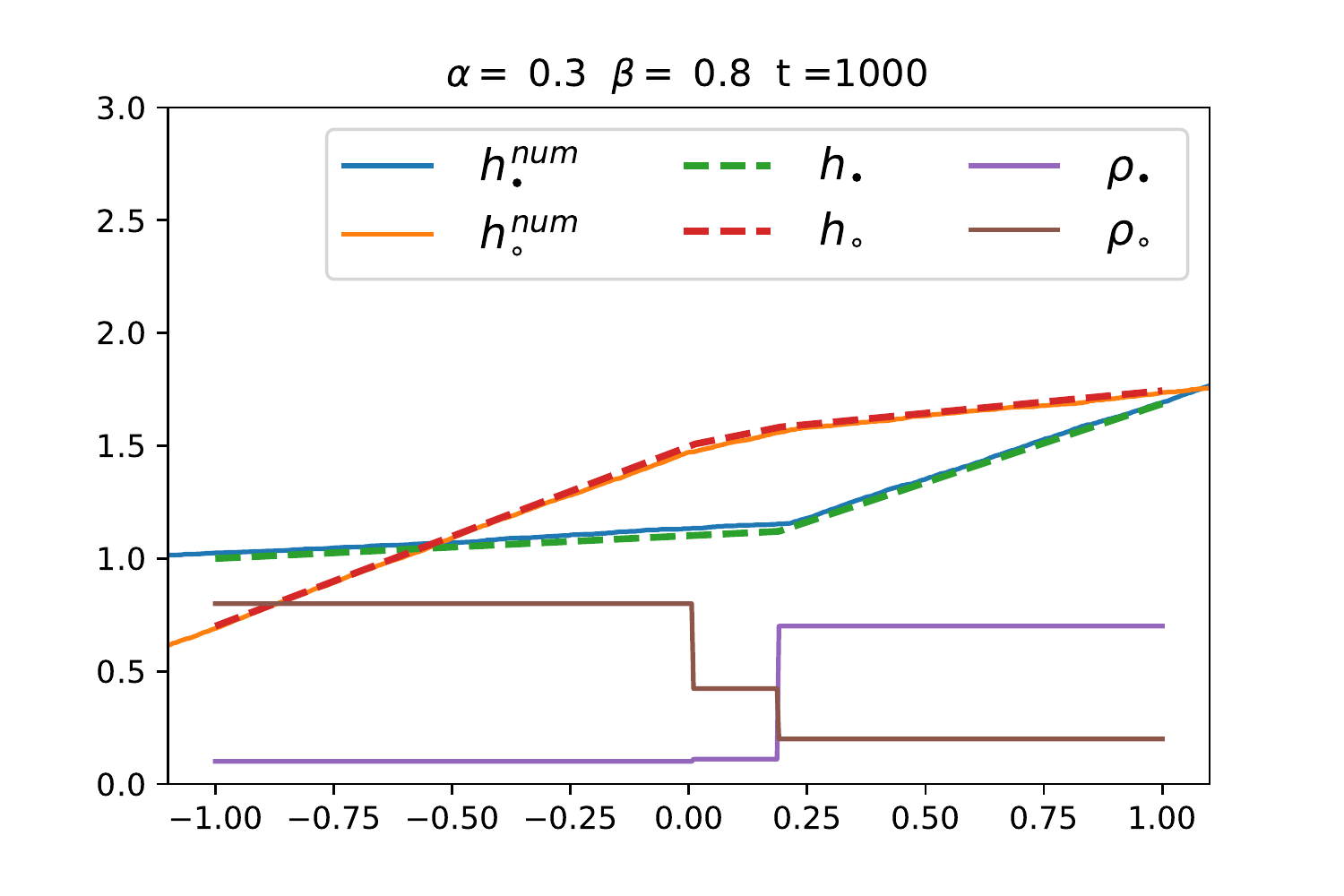}
		
		\label{fig:sub3}
	\end{subfigure}
	\begin{subfigure}{.44\textwidth}
		\centering
		
\begin{tikzpicture}

\begin{scope}[scale = .5]
\draw[thick,->] (0,0) -- (7,0) node[below] {\scriptsize $\rho_\circ$};
\draw[thick,->] (0,0) -- (0,7) node[left] {\scriptsize $\rho_\bullet$};

\foreach \rzL in {6*0.8}
\foreach \roL in {6*0.1}
\foreach \rzR in {6*0.2}
\foreach \roR in {6*0.7}
\foreach \rzi in {6*0.422}
\foreach \roi in {6*0.11}
{
	
	\draw [line width=0.7 mm,dashed, red, ->]  (\rzL,\roL)-- (\rzi*0.5 +\rzL*0.5 ,\roi*0.5 + \roL*0.5);
	\draw [line width=0.7 mm,dashed, red] (\rzi*0.5 +\rzL*0.5 ,\roi*0.5 + \roL*0.5) -- (\rzi,\roi);

	\draw [line width=0.7 mm ,red,dashed, ->]  (\rzi,\roi)-- (\rzi*0.5 +\rzR*0.5 ,\roi*0.5 + \roR*0.5);
	\draw [line width=0.7 mm,dashed, red] (\rzi*0.5 +\rzR*0.5 ,\roi*0.5 + \roR*0.5) -- (\rzR,\roR);
	
	\node[circle,fill=black,inner sep=0pt,minimum size=5pt,label=right:{\scriptsize $L$} ]  at (\rzL,\roL)  {};
	
	\node[circle,fill=black,inner sep=0pt,minimum size=5pt,label=west:{\scriptsize $R$} ]  at (\rzR,\roR) {};
	
	\node[circle,fill=black,inner sep=0pt,minimum size=5pt,label=south:{} ]  at (\rzi,\roi)  {};
	
}

\draw  (6*0.2,6-6*0.2) -- (6-0.4*6,0.4*6);
\draw (6*0.3,6-6*0.3) -- (0,6);
\draw (6,0) -- (6-0.4*6,0.4*6);

\draw [thick](2pt,6 cm) node[left] {\scriptsize $1$};
\draw [thick](6 cm,-2pt) node[below] {\scriptsize $1$};

\end{scope}

\end{tikzpicture}

		\label{fig:sub2}
	\end{subfigure}
\caption{Situation with two shocks}
\end{figure}
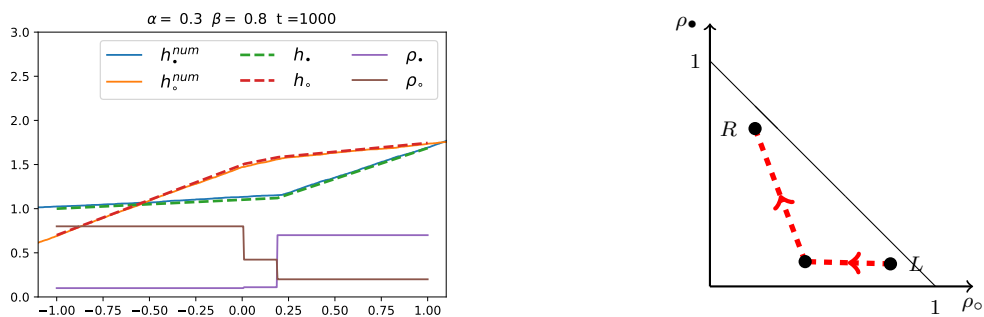

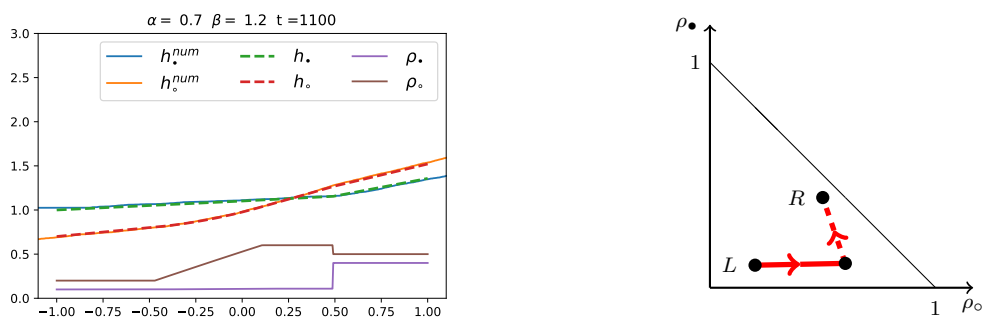
\begin{figure}[H]
	
		\begin{subfigure}{.55\textwidth}
		\centering
		
		\includegraphics[width=7cm]{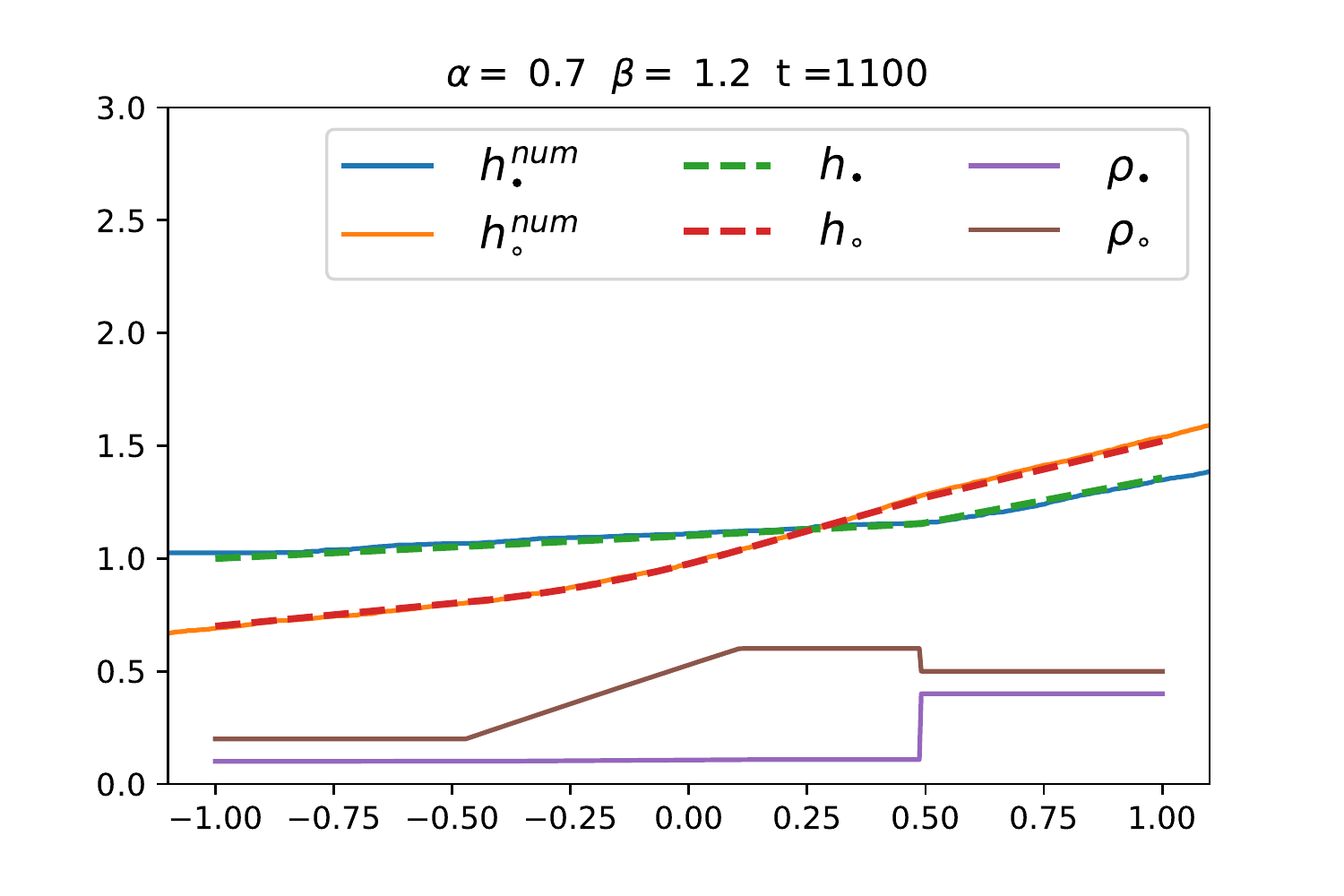}
		
		\label{fig:sub3}
	\end{subfigure}
	\begin{subfigure}{.44\textwidth}
		\centering
		
			\begin{tikzpicture}
		\begin{scope}[scale = .5]
		
		\draw[thick,->] (0,0) -- (7,0) node[below] {\scriptsize $\rho_\circ$};
		\draw[thick,->] (0,0) -- (0,7) node[left] {\scriptsize $\rho_\bullet$};
		
		\foreach \rzL in {6*0.2}
		\foreach \roL in {6*0.1}
		\foreach \rzR in {6*0.5}
		\foreach \roR in {6*0.4}
		\foreach \rzi in {6*0.600}
		\foreach \roi in {6*0.108}
		{
			
			\draw [line width=0.7 mm, red, ->]  (\rzL,\roL)-- (\rzi*0.5 +\rzL*0.5 ,\roi*0.5 + \roL*0.5);
			\draw [line width=0.7 mm, red] (\rzi*0.5 +\rzL*0.5 ,\roi*0.5 + \roL*0.5) -- (\rzi,\roi);

			\draw [line width=0.7 mm, dashed,red, ->]  (\rzi,\roi)-- (\rzi*0.5 +\rzR*0.5 ,\roi*0.5 + \roR*0.5);
			\draw [line width=0.7 mm,dashed, red] (\rzi*0.5 +\rzR*0.5 ,\roi*0.5 + \roR*0.5) -- (\rzR,\roR);
			
			\node[circle,fill=black,inner sep=0pt,minimum size=5pt,label=west:{\scriptsize $L$} ]  at (\rzL,\roL)  {};
			
			\node[circle,fill=black,inner sep=0pt,minimum size=5pt,label=west:{\scriptsize $R$} ]  at (\rzR,\roR) {};
			
			\node[circle,fill=black,inner sep=0pt,minimum size=5pt,label=west:{} ]  at (\rzi,\roi) {};
			
		}

		\draw  (6*0.2,6-6*0.2) -- (6-0.4*6,0.4*6);
		\draw (6*0.3,6-6*0.3) -- (0,6);
		\draw (6,0) -- (6-0.4*6,0.4*6);
		
		\draw [thick](2pt,6 cm) node[left] {\scriptsize $1$};
		\draw [thick](6 cm,-2pt) node[below] {\scriptsize $1$};
		\end{scope}
		\end{tikzpicture}

		\label{fig:sub2}

	\end{subfigure}

\caption{Situation with an $\alpha$--fan and a $\beta$--shock.}
\end{figure}

\begin{figure}[H]
	
		\begin{subfigure}{.55\textwidth}
		\centering
		
		\includegraphics[width=7cm]{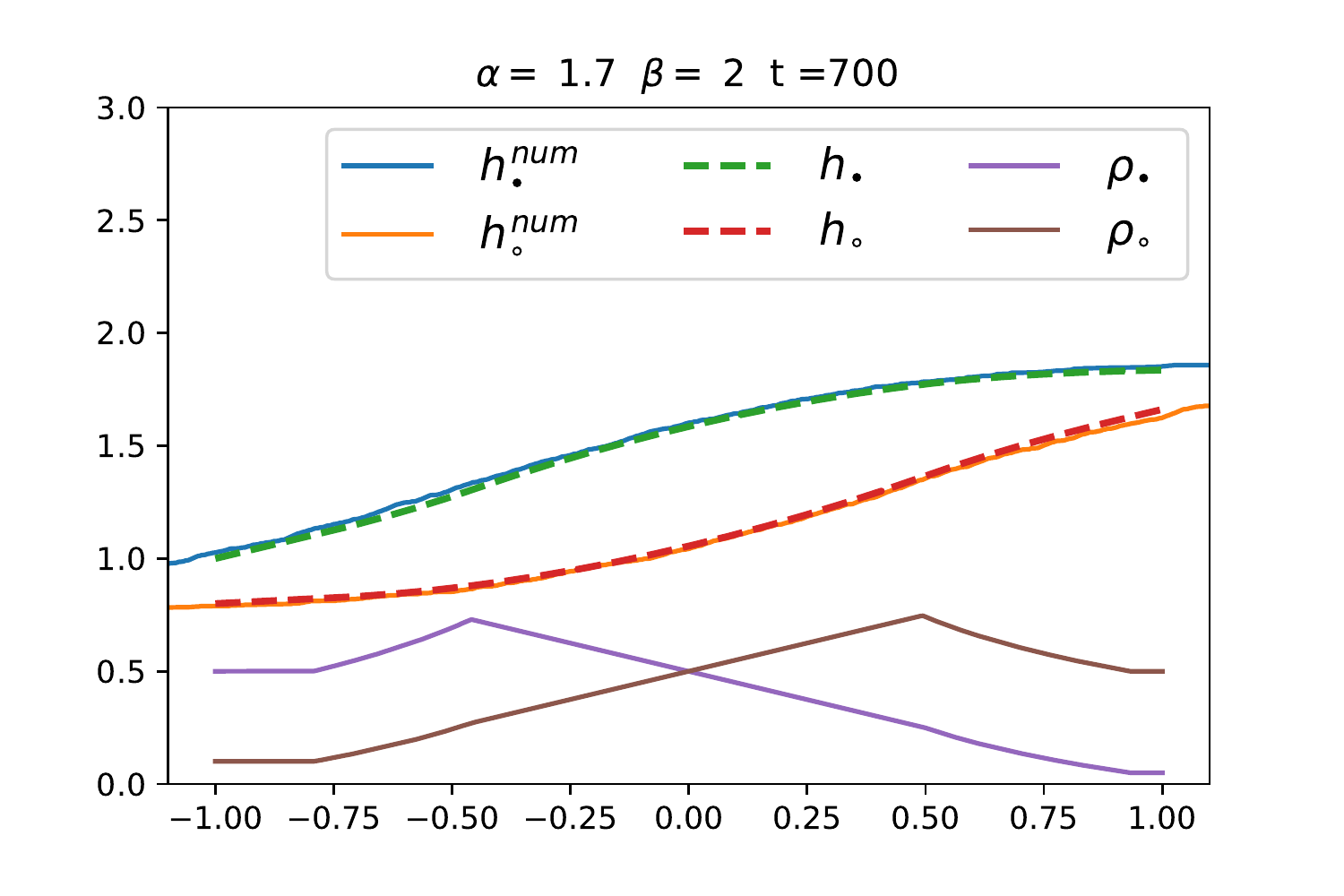}
		
		\label{fig:sub3}
	\end{subfigure}
   \begin{subfigure}{.44\textwidth}
		\centering
		
		\begin{tikzpicture}
		
		\begin{scope}[scale = .5]
\draw[thick,->] (0,0) -- (7,0) node[below] {\scriptsize $\rho_\circ$};
\draw[thick,->] (0,0) -- (0,7) node[left] {\scriptsize $\rho_\bullet$};

\foreach \rzL in {0.1*6}
\foreach \roL in {0.5*6}
\foreach \rzR in {0.5*6}
\foreach \roR in {0.05*6}
\foreach \rof in {6*0.729}
\foreach \rzf in {6*0.2701}
\foreach \ros in {6*0.2532}
\foreach \rzs in {6*0.7467}
{
	
	\draw [line width=0.7 mm, red, ->]  (\rzL,\roL) -- (\rzL*0.5 + \rzf*0.5, \rof*0.5 + \roL*0.5);
	
	\draw [line width=0.7 mm, red]  (\rzL*0.5 + \rzf*0.5, \rof*0.5 + \roL*0.5)-- (\rzf,\rof);

	\draw [line width=0.7 mm, red, ->]  (\rzf,\rof) -- (\rzf*0.5 + \rzs*0.5,\rof*0.5 + \ros*0.5);
	\draw [line width=0.7 mm, red] (\rzf*0.5 + \rzs*0.5,\rof*0.5 + \ros*0.5)  -- (\rzs,\ros);

	\draw [line width=0.7 mm, red, ->]  (\rzs,\ros) -- (\rzR*0.5+\rzs*0.5,\roR*0.5+\ros*0.5);
	\draw [line width=0.7 mm, red]  (\rzR*0.5+\rzs*0.5,\roR*0.5+\ros*0.5) -- (\rzR,\roR);

	\node[circle,fill=black,inner sep=0pt,minimum size=5pt,label=south:{\scriptsize $L$} ]  at (\rzL,\roL)  {};
	
	\node[circle,fill=black,inner sep=0pt,minimum size=5pt,label=west:{\scriptsize $R$} ]  at (\rzR,\roR) {};
	
}

\draw (6,0) -- (0,6);

\draw [thick](2pt,6 cm) node[left] {\scriptsize $1$};
\draw [thick](6 cm,-2pt) node[below] {\scriptsize $1$};
		\end{scope}
		
		\end{tikzpicture}

		\label{fig:sub2}
	\end{subfigure}
\caption{Situation diplaying an $\alpha$ and a $\beta$--fan with a depletion region in the middle. }
%
%
\end{figure}
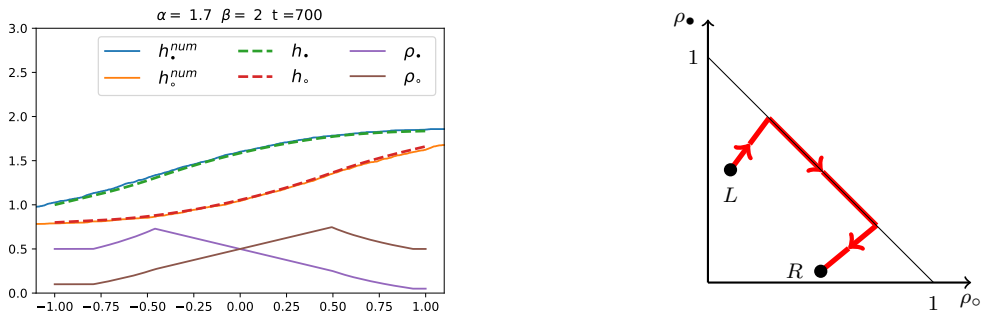

\section*{References}
\bibliographystyle{unsrt}
\bibliography{biblio-gen}

\end{document}